\documentclass[10pt]{article}

\usepackage{amsmath}
\usepackage{amsfonts}
\usepackage{amssymb}
\usepackage{amsthm}
\usepackage{latexsym}
\usepackage{listings}

\numberwithin{equation}{section}
\newtheorem{axm}{A}

\newtheorem{crule}{CR}
\newtheorem{definition}{Definition}

\newcommand{\be}{\begin{equation*}}
\newcommand{\ee}{\end{equation*}}
\newcommand{\bal}{\begin{aligned}}
\newcommand{\eal}{\end{aligned}}
\newcommand{\ba}{\begin{eqnarray}}
\newcommand{\ea}{\end{eqnarray}}

\newcommand{\mb}{\mathbb}

\newcommand{\lra}{\longrightarrow}
\newcommand{\lla}{\longleftarrow}

\begin{document} 

\title{Program Verification of Numerical Computation - Part 2.}

\author{Garry Pantelis}

\maketitle

\abstract

\noindent
These notes present some extensions of a formal method introduced in an earlier paper.
The formal method is designed as a tool for program verification of numerical computation and forms the basis of the software package \emph{VPC}.
Included in the extensions that are presented here are disjunctions and methods for detecting non-computable programs. 
A more comprehensive list of the construction rules as higher order constructs is also presented.

\section{Introduction}

In \cite{pan} a formal system for program verification of numerical computation was introduced and discussed in the context of the software package \emph{VPC}.
These notes outline some extensions of that formal system.
Most of the definitions and identities given in \cite{pan} will be repeated here for convenience and may contain some refinements.  
The important additions discussed in this paper include disjunctions and detecting non-computable programs.
Along with these additions is included an expanded collection of construction rules that will lay down the groundwork for analysis of the formal system that will presented elsewhere.

We are primarily interested in the computability of programs that rely heavily on numerical computation.
The task of establishing the computability of a program requires some kind of formal system based on inference.
In the search of a formal system that is suitable for our purposes one first looks to the methods of proof theory (see for example \cite{buss}).
The scope here is wide and one is immediately confronted by the need to make a choice of an inference scheme that is based on either classical or intuitionistic logic.
Since we are dealing with functional programs rather than formal statements in a more general sense one also needs to be guided by the topics of lambda calculus and in particular the typed lambda calculus (see for example \cite{baren}).

The formal methods in the general areas that have just been mentioned were primarily developed to address important theoretical questions and were not optimally designed for practical implementation in a machine environment. 
Apart from an issue of a translation of language there is no escaping the fact that regardless of the actual application for which any program has been constructed the computations are constrained by the finite resources of the machine on which the program is to be executed. 

We need to keep in mind that our objective is one of employing a classical computer to make inferences on programs that are to be executed on the same machine.
An important feature of a real world machine environment is the property of finite information storage along with a collection of well defined operations.
The context within which we have chosen to work is defined by the following machine specific parameters.
\be
\begin{array}{ll}
K & \text{number of characters in the alphabet.} \\
L & \text{maximum number of characters in any string.} \\
M & \text{maximum number of elements of any list stored as an array.} \\
\end{array}
\ee
The parameters $K,L$ and $M$ are dictated by the machine memory storage capacity and define the context within which our formal system is constrained.

Programs consist of a sequential order of statements that come in the form of instructions and assignments.
The sequential order is important although some rearrangement is allowed under special conditions.
To understand how this works we need to distinguish between variables that serve as input and output of a program.
Keeping track of the variable names is important in determining the constraints that allow one to manipulate statements that appear in a program.
Indeed it is this issue that largely motivates the construction of our formal system as an alternative to the conventional theories that are currently available.
More specifically, there is a need to define, in a more explicit way, the rules associated with the manipulation of statements that come in the form of functional programs that are defined in terms of the variable names associated with their input/output lists.
With this in mind one should not lose sight of the constraints imposed by the finite resources of the machine on which the programs are to be executed.

In a program setting we deal with objects and types. Types will be denoted by the symbols $\mb{A},\mb{B},\ldots,\mb{Z}$.
Object $a$ has type $\mb{A}$ is denoted by $a:\mb{A}$. An object may also be dependent on another object.
We write $a(n)$ to mean that the object $a$ depends on the parameter or object $n$.

Types may be subtypes of types.
$\mb{A}$ is a subtype of $\mb{B}$ is denoted by $\mb{A}<:\mb{B}$.
Types may also be dependent on objects.
We write $\mb{A}(a_1,\ldots,a_n)$ to mean that type $\mb{A}$ objects depend on the parameters or objects $a_1,\ldots,a_n$.
Parameter dependent types are subtypes of their generic type, i.e. $\mb{A}(a_1,\ldots,a_n)<:\mb{A}$. 
It is worth noting that if $A <: B$ and $\mb{A}(a_1,\ldots,a_n)$ and $\mb{B}(b_1,\ldots,b_m)$ are parameter dependent subtypes it does not necessarily follow that $\mb{A}(a_1,\ldots,a_n)$ is a subtype of $\mb{B}(b_1,\ldots,b_m)$.

Programs are written in a language that employ characters of an alphabet.
The alphabet we will work with consists of lower and upper case letters ($a-z$, $A-Z$), digits (0-9) and the special characters
\be
~.~,~+~-~*~/~(~)~[~]~|
\ee
These are the same special characters used in \cite{pan} with the addition of the character '$|$' that will be needed for disjunctions.

A string of the alphabet is a sequence of characters $s(i_1) s(i_2) \ldots s(i_j)$, $1 \leq i_1,\ldots,i_j \leq K$, $1 \leq j \leq L$, where each $s$ is a character of the alphabet.
A string is given a type denoted by $\mb{S}$.
There are two main subtypes of strings that will be considered here.

\be
\begin{array}{ll} 
\mb{C}<:\mb{S} & \text{alphanumeric strings comprised of any combination of letters and digits} \\
& \text{with the first character always being a letter.} \\
 \mb{I}<:\mb{S} & \text{signed integers comprised of digits preceeded by a sign $\pm$.} \\
\end{array}
\ee

Alphanumeric strings are assigned the type $\mb{C}$ and are often used to represent names of programs and variable names of elements of input/output (I/O) lists of programs.
Variable names of elements of I/O lists of programs serve as place holders for assigned values that are defined as specific types within the program.
We write $a:\mb{C}$ to stress that $a$ is a dummy variable that represents an alphanumeric string.
Upon entry to a program we may also write $a:\mb{A}$ to denote that the alphanumeric string has been assigned a value of type $\mb{A}$.
The assigned value can be any string of a specific subtype.
 
If $a$ and $b$ are dummy variables representing two strings we write $a=b$ to denote that the two strings are identical.
The sense in which the equality is being used here will always be assumed unless otherwise stated.

We may also write $a=b$ to mean that the assigned value of the alphanumeric variable represented by the dummy variable $a$ is identical to the assigned value of the alphanumeric variable represented by the dummy variable $b$.
The sense in which the equality is used here will always be stated to avoid confusion.
For two alphanumeric variable names represented by $a$ and $b$ we write $a:=f(b)$ to denote that the value assigned to $a$ is acquired through the assignment function $f$ acting on the value assigned to $b$.

An object of type $\mb{I}$ is a string that can be assigned any one of the integer values
\be
0, \pm 1,\ldots, \pm N ,
\ee
where $N$ is the maximum positive integer and is a specific machine parameter.
We write $a:\mb{I}_0$ to mean that $a:\mb{I}$ and $0 \leq a \leq N$ and $a:\mb{I}_+$ to mean that $a:\mb{I}$ and $0 < a \leq N$.
We adopt the usual convention of dropping the prefix $+$ sign when dealing with positive integers.
 
A list has the representation $a=[a_1,\ldots,a_m]=[a_i]_{i=1}^m$, for some $m:\mb{I}_0$, where each element $a_i$, $i=1,\ldots,m$, is an object of some specified type.
The case $m=0$ represents the empty list $[~]$.
Elements of lists may themselves be lists.
We write $a:\mb{L}$ to mean that $a$ is an object of type list with unspecified length.
We may also write $a:\mb{L}(m)$, for some $m:\mb{I}_0$, to mean that $a$ is an object of type list with specified length $m$.
The concatenation $c$ of two lists $a=[a_i]_{i=1}^m$ and $b=[b_i]_{i=1}^n$ is the list given by
\be
c=[a,b] = \big [ [a_1,\ldots,a_m],[b_1,\ldots,b_n] \big ] = [a_1,\ldots,a_m,b_1,\ldots,b_n]
\ee
Here the internal brackets that delimit $a$ and $b$ can be removed so that $c$ becomes an object of type $\mb{L}(m+n)$.
The above will be referred to as the standard list concatenation.
Standard list concatenation does not always apply to program I/O lists where $c=[a,b]$ can be regarded as an object of type $\mb{L}(2)$ with two elements of type $\mb{L}(m)$ and $\mb{L}(n)$.
These rules can be extended in the obvious way for concatenations involving more than two lists. 

A list $b:\mb{L}(m)$, $m:\mb{I}_+$, is a sublist of $a=[a_1,\ldots,a_n]$, $m \leq n$, if it has the representation $b=[a_{i_1},\ldots,a_{i_m}]$, where $[i_1,\dots,i_m]$ is as an $m$-permutation of $[1,\ldots,n]$.
We write $b \subseteqq a$ to mean that $b$ is a sublist of $a$.
When $m<n$ we say that $b$ is a strict sublist of $a$ and write $b \subset a$.
When $m=n$ we either have $b=a$ or $b$ is a permutation of $a$. In either case we also have $a \subseteqq b$. 
The empty list $[~]$ is regarded to be a sublist of all lists.

\section{Programs.}

\vspace{5mm}
\noindent
\textbf{Program structure.}
Programs are strings with a specific structure and are assigned the type denoted by $\mb{P} <: \mb{S}$.
Program names are assigned the type $\mb{P}_{name}$ and are a specific subtype of alphanumeric strings, i.e. $\mb{P}_{name} <: \mb{C}$.
The convention used here is to represent all program names in the form of a string as a combination of letters and digits, with the starting character being an upper case letter and any following letters in lowercase. Programs are defined recursively as follows.

\begin{definition}(Program.)
A program has the representation $P(x,y)$ with the allocation of types of its component parts given by
\be
\begin{array}{ll}
P(x,y):\mb{P} & \text{program} \\
P:\mb{P}_{name} & \text{program name} \\
x:\mb{L} & \text{input list} \\
y:\mb{L} & \text{output list} \\
\end{array}
\ee

A program $P(x,y)$ satisfies all of the following conditions.

\begin{itemize}

\item Elements of the I/O lists of a program are alphanumeric variable names (type $\mb{C}$) that serve as placeholders for assigned values.
The type of the assigned value of every element of the I/O lists are checked within the program.

\item The variable names of the elements of the output list $y$ are distinct.

\item No element of the input list $x$ can have a variable name that coincides with a variable name of an element of the output list $y$.

\end{itemize}

\noindent
A program $P(x,y):\mb{P}$ can be represented by a list $P(x,y) = [P_i(x_i,y_i)]_{i=1}^n$, for some $n:\mb{I}_+$, where $P(x,y)$ is referred to as the main program and each $P_i(x_i,y_i):\mb{P}$, $i=1,\ldots,n$, is referred to as a subprogram of $P(x,y)$.
The following convention is adopted that defines the I/O lists of the main program with respect to the I/O lists ($x_i$, $y_i$, $i=1, \ldots ,n$) of its subprograms.
 
\begin{itemize}

\item $y = [ y_i ]_{i=1}^n$, where $[y_i]_{i=1}^n$ is the standard list concatenation of the lists $y_1,\ldots,y_n$.

\item $x = x^\prime \setminus (x^\prime \bigcap y)$, where $x^\prime \simeq [x_i]_{i=1}^n$ and $[x_i]_{i=1}^n$ is the standard list concatenation of the lists $x_1,\ldots,x_n$.

\end{itemize}

\noindent
A program list satisfies all of the following conditions.

\begin{itemize}

\item $\bigcap_{i=1}^n y_i = [~]$.

\item $x_i~\bigcap~[y_j]_{j=i}^n = [~]$, $i=1,\ldots,n$, where $[y_j]_{j=i}^n$ is a standard list concatenation.

\item $P \neq P_i$, $i=1,\ldots,n$, for $n>1$.

\end{itemize}
The empty program is denoted by $[~]$.

\end{definition}

\noindent
Note that $x_i$ and $y_i$ are lists in their own right and are not necessarily meant to represent elements of any particular list.   
The statement $x^\prime \simeq [x_i]_{i=1}^n$ means that $x^\prime$ is the standard list concatenation of the lists $x_1,\ldots,x_n$, with repeated variable names removed by following the procedure outlined in \cite{pan}.
The symbols $\setminus$ and $\bigcap$, respectively, denote list minus and list intersection, respectively, analogous to the way that they are used in set theory.

\vspace{5mm}
\noindent
\textbf{Computability.} 
Within all programs type checking is performed on the assigned values of all elements of its I/O lists.
Execution errors are associated with type violations.
A program will halt with an execution error if during its execution there is a type violation of any assigned value of the elements of its I/O lists.
The execution of a program list is completed when the execution of all subprograms of the list have been successfully completed in the sequential order from left to right.  

\begin{definition}(Computability.)
A program $P(x,y):\mb{P}$ is said to be computable, with respect to the values assigned to its input list $x$, if upon execution it halts without encountering an execution error.
A computable program returns the output $y$, where $y$ may be the empty list.
We write $P(x,y):\mb{P}_{comp}$ to mean that the program $P(x,y)$ is computable in the sense that it is computable for at least one value assigned input list $x$.
We write $P(x,y):\mb{P}_{comp}(x)$ to stress that $P(x,y)$ is computable for the specific value assigned input list $x$.
\end{definition}

\begin{definition}(False program.)
A program $P(x,y):\mb{P}$ is a false program if there does not exist any value assigned input list $x$ such that $P(x,y)$ is computable.
A false program $P(x,y)$ is assigned the type $P(x,y):\mb{P}_{false}$ with the hierarchy of subtypes $\mb{P}_{false}<:\mb{P}$.
\end{definition}

\vspace{5mm}
\noindent
\textbf{Atomic programs.}
Programs are constructed from lists of atomic functional programs.
Atomic programs are given the type $\mb{P}_{atom} <: \mb{P}$ and can be partitioned into the following subtypes.

\vspace{5mm}
\noindent
$\mb{P}_{type}$. Atomic programs of type $\mb{P}_{type}$ have the sole task of checking the type of the assigned value of each element of its input list.
If there is a type violation the program will halt with an execution error.
$\mb{P}_{type}$ programs have an empty list output and have the general structure $P(x,[~])$.

\vspace{3mm}
\noindent
$\mb{P}_{assign}$. Atomic programs of type $\mb{P}_{assign}$ assign values to their output variables through some assignment function acting on the assigned values of their input variables.
Within each $\mb{P}_{assign}$ program type checking is performed on the values assigned to all elements of its I/O lists.
If there is a type violation the program will halt with an execution error. 

\vspace{3mm}
\noindent
$\mb{P}_{tassign}$. Objects acquire their type either by default or are assigned.
Atomic programs of type $\mb{P}_{tassign}$ assign a subtype to a structure involving one or more elements of its input list.
Upon entry the type of the assigned values of the elements of the input list are checked (not the subtype that is to be assigned).
If there is a type violation the program will halt with an execution error.
Type $\mb{P}_{tassign}$ programs have an empty list output and have the general structure $P(x,[~])$.

\vspace{5mm}
\noindent
\textbf{Program equivalence.}
Program equivalence refers to programs that may appear to have a different structure but are functionally identical.
Program equivalence will be defined in terms of the sequential ordering of the program list.
This definition will be generalized later.

\begin{definition}(Program equivalence.)
Two programs given by the list representations $[P_i(x_i,y_i)]_{i=1}^n$ and $[P_i^\prime(x_i^\prime,y_i^\prime)]_{i=1}^n$ are said to be program equivalent provided that $[P_i^\prime(x_i^\prime,y_i^\prime)]_{i=1}^n$ is a list permutation of $[P_i(x_i,y_i)]_{i=1}^n$.
Program equivalence is denoted by $[P_i(x_i,y_i)]_{i=1}^n \equiv [P_i^\prime(x_i^\prime,y_i^\prime)]_{i=1}^n$ and satisfies the properties of reflexivity, symmetry and transitivity

\end{definition}

The second important type of equivalence involves programs that differ in variable names used to define their I/O lists but can be associated with some degree of functionality.
Before giving the definition for I/O equivalence we need to make a distinction between common variables and constants.
For each type there may exist specific objects of that type that are of particular interest because they may appear as fixed input parameters to a program.
For example the type integers, $\mb{I}$, has three constants $-1,0,1$.
For higher order programs, where programs themselves serve as inputs, we may regard the empty program $[~]$ as a constant for type $\mb{P}$.

\begin{definition}(I/O equivalence.)
Consider two programs with the list representations $[P_i(x_i,y_i)]_{i=1}^n$ and $[P_i(x_i^\prime,y_i^\prime)]_{i=1}^n$.
For each $i=1,\ldots,n$, let $m_i$ be the length of the lists $x_i$ and $x_i^\prime$ and denote by $x_{ik}$ and $x_{ik}^\prime$, respectively, the $k$th element of $x_i$ and $x_i^\prime$, respectively.
Similarly, for each $i=1,\ldots,n$, let $n_i$ be the length of the lists $y_i$ and $y_i^\prime$ and denote by $y_{ik}$ and $y_{ik}^\prime$, respectively, the $k$th element of $y_i$ and $y_i^\prime$, respectively.
The program $[P_i(x_i^\prime,y_i^\prime)]_{i=1}^n$ is I/O equivalent to the program $[P_i(x_i,y_i)]_{i=1}^n$ provided that all of the following conditions are satisfied.

\begin{itemize}

\item If $x_{ik} = x_{jl}$ then $x_{ik}^\prime = x_{jl}^\prime$, $1 \leq i \leq n$, $1 \leq k \leq m_i$, $1 \leq j \leq i$, $1 \leq l \leq m_j$.

\item If $x_{ik} = y_{jl}$ then $x_{ik}^\prime = y_{jl}^\prime$, $2 \leq i \leq n$, $1 \leq k \leq m_i$, $1 \leq j \leq i-1$, $1 \leq l \leq n_j$.
\item If $x_{ik}$ is a constant then $x_{ik}^\prime$ is also the same constant, $1 \leq i \leq n$, $1 \leq k \leq m_i$.

\end{itemize}
$[P_i(x_i^\prime,y_i^\prime)]_{i=1}^n$ is I/O equivalent to the program $[P_i(x_i,y_i)]_{i=1}^n$ is denoted by $[P_i(x_i^\prime,y_i^\prime)]_{i=1}^n \thicksim [P_i(x_i,y_i)]_{i=1}^n$.
I/O equivalence does not satisfy the property of symmetry.

\end{definition}

\vspace{5mm}
\noindent
\textbf{Notes.}

\begin{itemize}

\item The empty list program is denoted by $[~]$ and is always computable.
If the empty list program is an element of a program list it can be extracted from the program list without effecting the functionality of that program.

\item In the definition of a false program it is stated that $\mb{P}_{false}<:\mb{P}$.
This means that for an object to have the type assignment $\mb{P}_{false}$ it must first have the structure of a program under the formal definition of a program.
The statement that a program will always halt as a result of a syntactic error is not considered meaningful in this context since such an object cannot be assigned the type $\mb{P}$. 

\end{itemize}

\section{The Sublist Rule.}

Programs are strings with a particular structure and may serve as assigned values of elements of an I/O list of a program.
A program will be said to be a higher order program if the assigned values of the elements of its I/O lists are of type $\mb{P}$.
Higher order programs essentially recognize the structure of programs as strings of a particular subtype. 
Here we state the program construction rules as constructs of higher order programs.

We use the shorthand notation of representing programs with lower case letters so that, for example, $a:\mb{P}$ is understood to be a dummy variable that represents a string of subtype $\mb{P}$.
We may also regard $a$ as being assigned the value of type $\mb{P}$ explicitly given by the program $P_a(x_a,y_a)$.

\begin{definition}(Computable program extension.)
A program $s:\mb{P}$ has the subtype $s:\mb{P}_{cpe}$ if it admits a decomposition $s=[p,c]$ such that if $p$ is computable with respect to a value assigned input then $s$ is also computable for the same value assigned input.
The hierarchy of subtypes is $\mb{P}_{cpe}<:\mb{P}$.
The program $s=[p,c]$ is said to be a computable program extension of the program $p$.
We write $s:\mb{P}_{cpe}(p,c)$ to stress that $s$ has type $\mb{P}_{cpe}$ under the decomposition $s=[p,c]$.
\end{definition}

\begin{definition}(Irreducible computable program extension.)
A program $s:\mb{P}$ is called an irreducible computable program extension and assigned the subtype $s:\mb{P}_{icpe}$ if it admits a decomposition $s=[p,c]$ such that all of the following conditions are satisfied.

\begin{itemize}

\item $s:\mb{P}_{cpe}(p,c)$.

\item The program $p$ is irreducible in the following sense. There does not exist a program $r=[q,c]$ such that $q \subset p$ ($q$ is a strict sublist of $p$) and $r:\mb{P}_{cpe}(q,c)$.

\end{itemize}
The hierarchy of subtypes is $\mb{P}_{icpe}<:\mb{P}_{cpe}<:\mb{P}$.
The programs $p$ and $c$, respectively, are said to be the premise and conclusion, respectively, of the irreducible computable program extension $s$.
We write $s:\mb{P}_{icpe}(p,c)$ to stress that $s$ has type $\mb{P}_{icpe}$ under the decomposition $s=[p,c]$.
\end{definition}

Higher order programs will be constructed from the atomic programs whose names are given in the tables below.
All atomic programs are defined in the appendix.
 
\vspace{5mm}

\begin{tabular}{|c|c|}
\hline
Atomic program names & Atomic program type \\
\hline
$Prog,~Equiv,~Eqio,~Sub,~Cpe,~False$ & $\mb{P}_{type}$ \\
\hline
$Conc,~Disj$ & $\mb{P}_{assign}$ \\
\hline
$Acpe,~Afalse$ & $\mb{P}_{tassign}$ \\
\hline
\end{tabular} 

\vspace{5mm}

\begin{definition}(Sublist derivation.)
A sublist derivation $s:\mb{P}$ with respect to the program $[q,c]:\mb{P}$ is an assignment $s:=[p,c]$ subject to the conditions $q \subseteqq p$, $[q,c]:\mb{P}_{cpe}(q,c)$ and $[p,c]:\mb{P}$.
It is constructed from the higher order program $Sd([q,p,c],[s])$ defined by
\be
Sd([q,p,c],[s]) = \big [ Sub([q,p],[~]),~Cpe([q,c],[~]),~Conc([p,c],[s]) \big ]
\ee
$Sd([q,p,c],[s])$ is called the sublist derivation program.
\end{definition}

\vspace{5mm}
\noindent
\textbf{Construction rules.}
The following construction rules are presented as irreducible computable program extensions of higher order constructs.
The internal square brackets delimit the premise program from the conclusion.
(The standard list concatenation for programs apply so that the internal brackets can be removed.)
When the premise program contains only a single statement the internal square brackets are omitted.
The main inference rule is the sublist rule.

\vspace{5mm}
\noindent
Sublist rule.
\begin{crule}\label{slr}
\be
\bal
& \Big [ Sd([q,p,c],[s]),~Acpe([p,c],[~]) \Big ] \\
\eal
\ee
\end{crule}

The two rules that follow involve the acquisition of the property of a computable program extension through program and I/O equivalence.
The final rule states that once assigned, the property of a computable program extension is retained.
In other words, once a program has been assigned the type $\mb{P}_{cpe}$ it is stored in memory as such so that it retains that type whenever it is accessed by any following subprogram of a higher order program list. 

\vspace{5mm}
\noindent
Equivalence of computable program extensions.
\begin{crule}\label{eqvcpe}
\be
\bal
& \Big [ \big [ Cpe([p,c],[~]),~Equiv([p,q],[~]),~Equiv([c,d],[~]) \big ],~Acpe([q,d],[~]) \Big ] \\
\eal
\ee
\end{crule}

\noindent
I/O equivalence of computable program extensions.
\begin{crule}\label{ioequiv}
\be
\bal
& \Big [ \big [ Cpe([q,c],[~]),~Conc([q,c],[r]),~Conc([p,d],[s]),~Eqio([p,q],[~]),~Eqio([s,r],[~]) \big ],~Acpe([p,d],[~]) \Big ] \\
\eal
\ee
\end{crule}

\noindent
Retention of subtype assignment. 
\begin{crule}\label{rsa}
\be
\bal
& \big [ Acpe([p,c],[~]),~Cpe([p,c],[~]) \big ] \\
\eal
\ee
\end{crule}

\vspace{5mm}
\noindent
\textbf{Derivations and proofs.}
A program $[p_i]_{i=1}^m$ is called a derivation if it is constructed by a sequence of sublist derivations.
Let $s_k=[p_i]_{i=1}^k$ and consider the following iteration.

\begin{itemize}

\item The program $s_n=[p_i]_{i=1}^n$, for some $n<m$, serves as a list of premises of the derivation.

\item For each iteration $i=n+1,\ldots,m$, the statement $p_i$, is obtained from a sublist derivation $Sd([q_i,s_{i-1},p_i],[s_i])$, for some $[q_i,p_i]:\mb{P}_{cpe}(q_i,p_i)$.

\end{itemize}

\noindent
A derivation may be called a proof if its final statement is of particular interest in relation to its premise program.
Irreducible computable program extensions that are extracted from proofs are called theorems.
Axioms are irreducible computable program extensions for which no derivation is known.

\vspace{5mm}
\noindent
\textbf{Notes.}

\begin{itemize}

\item The sublist rule states that if $s=[p,c]$ is a sublist derivation with respect to the computable program extension $[q,c]$ then it follows that $s$ is a computable program extension of $p$.
The conclusion program of the sublist rule is a type assignment $[p,c]::\mb{P}_{cpe}(p,c)$.

\item In \cite{pan} program equivalence is based solely on sequential equivalence.
Here we have replaced the atomic program name $Eqseq$ used in \cite{pan} with the atomic program name $Equiv$.
We do this because later we will extend the notion of program equivalence to include other program structures.
Also there is an additional statement of program equivalence in CR2 associated with the conclusion program $c$.
This generalizes the previous version of CR2 for the reasons discussed in the following note.

\item In a statement of the form $[p,c]:\mb{P}_{cpe}(p,c)$ the parameter dependence $(p,c)$ of the type $\mb{P}_{cpe}$ may appear superfluous.
This is not the case because the programs $p$ and $c$ may be program lists and hence the program $[p,c]$ may be expressed in terms of other decompositions of the form $[p^\prime,c^\prime]$.
The statement $\mb{P}_{cpe}(p,c)$ stresses that the program $[p,c]$ is assigned the type $\mb{P}_{cpe}$ under the specific decomposition $[p,c]$.

\item We have also changed the definition of the programs $Cpe$ and $Acpe$ as given in \cite{pan} by removing the third variable in the input list.
The program $Cpe([p,c],[~])$ is a type checking program that checks $p:\mb{P}$, $c:\mb{P}$ and $[p,c]:\mb{P}_{cpe}(p,c)$.
An assignment $s:=[p,c]$ can be made by applying CR15 given in a later section.
As a result of these changes the fourth input parameter that appeared in the original version of the program $Sd$ is now redundant and does not appear in the input list of the current version. 

\item In \emph{VPC}, the program $[q,c]$ of a sublist derivation $Sd([q,p,c],[s])$ is an axiom or theorem although in the sublist rule it suffices that $[q,c]$ be a computable program extension.

\item During proof construction, \emph{VPC} accesses a file \emph{axiom.dat} that initially stores all of the axioms of the specific theory under investigation.
As proofs are completed the theorems extracted from them are also stored in \emph{axiom.dat}.

\item Axioms and theorems stored in the file \emph{axiom.dat} are assigned the type $\mb{P}_{cpe}$ by default.
Otherwise a program acquires the type $\mb{P}_{cpe}$ through the type assignment program $Acpe$.

\item In a sublist derivation $Sd([q,p,c],[s])$, the program $[q,c]$ is identified as an axiom/theorem if there exists an axiom/theorem $[q^\prime,c^\prime]$ stored in the file \emph{axiom.dat} and there exists programs $q^{\prime \prime} \equiv q$ and $c^{\prime \prime} \equiv c$ such that $q^{\prime \prime}$ is I/O equivalent to $q^\prime$ and $[q^{\prime \prime},c^{\prime \prime}]$ is I/O equivalent to $[q^\prime,c^\prime]$.

\item An important feature of \emph{VPC} is that at each iteration of a derivation all possible sublist derivations of the current program are extracted.
The conclusions to each sublist derivation are written to a file \emph{options.dat} that can be accessed by the user.
The user selects one of these conclusions to generate the next statement in a proof.
The procedure is one of extracting all sublists of the current program that can be matched to the premise programs of the axioms/theorems stored in \emph{axiom.dat}.
The matching is based on the rules of program and I/O equivalence of computable program extensions as described in the previous note.
\emph{VPC} employs special techniques to short cut the matching procedures so that computations do not become excessive. 

\item A sublist derivation $Sd([q,p,c],[s])$ is not in itself a computable program extension, i.e. the last statement $Conc([p,c],[s])$ is not necessarily computable if the first two statements are computable.
Thus, a necessary condition that $Sd([q,p,c],[s])$ be computable is that $[p,c]:\mb{P}$.
The concatenation $[p,c]$ may fail if for instance the two conditions $y_c \bigcap x_p = [~]$ and $y_c \bigcap y_p = [~]$ are not satisfied.
It is often the case that one is able to ensure in advance that $[p,c]:\mb{P}$ by a suitable choice of the variable names of the elements of the output list, $y_c$, of $c$.

\item Derivations and proofs have the further distinguishing feature in that they come with a collection of connection lists.
Consider the derivation $P(x,y)=[P(x_i,y_i)]_{i=1}^m$.
For each subprogram, $P_i(x_i,y_i)$, that is not a premise, is associated with the connection list of the form
\be
\big [ A(i),l(i,1),\ldots,l(i,k(i)) \big ]
\ee
where $1 \leq l(i,1),\ldots,l(i,k(i)) \leq i-1$ are the labels of the subprograms of the sublist of $P(x,y)$ that coincides with the premise program of the axiom/theorem, labeled $A(i)$, that is used to conclude $P_i(x_i,y_i)$.
Here $k(i)$ is the length of the premise program list of the axiom/theorem $A(i)$.

\end{itemize}

\section{Axioms/theorems of falsity}

While a program $p:\mb{P}_{false}$ will halt with an execution error for any value assigned input it does not necessarily follow that the sublist derivation program $Sd([q,p,c],[s])$ will also halt with an execution error.
The reason for this is that $Sd([q,p,c],[s])$ is a higher order program so that type checking is based on program structure.
It does not recognize value assignments of the I/O lists of the programs $q,p,c$ and $s$.
Thus there is nothing stopping us from allowing the program $p$ of a sublist derivation $Sd([q,p,c],[s])$ to be of type $\mb{P}_{false}$.
If $Sd([q,p,c],[s])$ does not halt with an execution error the derived object $s$ will be a program but will also be of subtype $\mb{P}_{false}$.
In this section we demonstrate how sublist derivations can be used to identify false programs.

The statement $p:\mb{P}_{false}$ may be represented by the higher order construct $False([p],[~])$.
If $p$ is irreducible in the sense that there does not exist a program $q \subset p$ ($q$ is a strict sublist of $p$) such that $q:\mb{P}_{false}$ then the statement $False([p],[~])$ represents an axiom or a theorem of falsity.
Axioms and theorems of falsity are higher order constructs of irreducible computable program extensions with an empty list premise.
 
To the construction rules we introduce the additional rules of falsity.

\vspace{5mm}
\noindent
Sublist falsity rule.

\begin{crule}\label{sfr}
\be
\bal
& \Big [ \big [ Sub([q,p],[~]),~False([q],[~]) \big ],~Afalse([p],[~]) \Big ] \\
\eal
\ee
\end{crule}

\vspace{5mm}
\noindent
Equivalence of false programs.

\begin{crule}\label{efp}
\be
\bal
& \Big [ \big [ False([p],[~]),~Equiv([p,q],[~]) \big ],~Afalse([q],[~]) \big ] \\
\eal
\ee
\end{crule}

\vspace{5mm}
\noindent
Retention of subtype assignment.

\begin{crule}\label{rsa2}
\be
\bal
& \big [ Afalse([p],[~]),~False([p],[~]) \big ] \\
\eal
\ee
\end{crule}

Consider a premise program $p$, where $p=[p_i]_{i=1}^n$.
Since we have allowed that the premise program $p$ to be of type $\mb{P}_{false}$ we may iteratively apply the sublist derivation program $Sd$ to generate the program $[p,q]$, where $q=[q_i]_{i=1}^m$ are the statements obtained by $m$ sublist derivations.
If $p:\mb{P}_{false}$ the iteration should continue until a sublist of $[p,q]$ coincides with a false program defined in an axiom or theorem of falsity stored in \emph{axiom.dat}.
When this occurs we abandon the sublist derivation format and apply the sublist falsity rule.

In \emph{VPC} the output of the proof program will look like the following vertical list (the connection lists are omitted).

\be
\begin{array}{ll}
Label  & Statement \\
1      & p_1       \\
\vdots & \vdots    \\
n      & p_n        \\
n+1      & q_1       \\
\vdots & \vdots    \\
n+m     & q_m        \\
n+m+1     & False   \\
\end{array}
\ee  
The first $n+m$ lines are in the standard derived proof format.
The addition of the final statement $False$ means that the standard proof format is to be abandoned and the vertical list is to be read as the statement
\be
[p,q]:\mb{P}_{false}
\ee
We may extract from this statement a theorem of falsity $False([p],[~])$, provided that $p$ is minimal in the sense that there are no strict sublists of $p$ that are of type $\mb{P}_{false}$.

\vspace{5mm}
\noindent
\textbf{Notes.}

\begin{itemize}

\item In the derivation $[p,q]$ depicted in the above table the connection lists for the derived statements of lines $n+1$ to $n+m+1$ have been omitted.
The connection lists can be used to trace back the falsity of the premise $p$ from the derivation $[p,q]:\mb{P}_{false}$.
Extraction of theorems of falsity follow the same algorithm for theorem extraction as described in \cite{pan}.  

\item Given that the premise program of a sublist derivation could be of type $\mb{P}_{false}$ one should avoid terminating an iteration of derivations before a conclusion leads to a statement of falsity. 
This is to prevent an intermediary derivation being identified as a proof of a theorem.
More will be said on this later.

\item There is an important consequence of allowing the program $p$ of a sublist derivation $Sd([q,p,c],[s])$ to be of type $\mb{P}_{false}$.
If we accept the sublist rule, CR1, without exception we must conclude that there exist computable program extensions $[p,c]$ such that $p:\mb{P}_{false}$.
Careful reading of the definition for computable program extensions does not disallow such a possibility.
The definition states that if the premise $p$ is computable for a given value assigned input then it is guaranteed that $[p,c]$ is computable for the same value assigned input.

\end{itemize}

\section{Disjunctions.}

In conventional theories of logic, disjunctions have an important role to play in the expressiveness of formal statements.
Program disjunctions have a more basic role in that they effectively split a program into several parallel programs, where each parallel program is associated with an operand of the disjunction contained within the main program.
Once a disjunction splitting has been completed sublist derivations can be performed independently on each parallel program.

\begin{definition}(Disjunction.)
A program $P(x,y):\mb{P}$ is a disjunction if it has the form
\be
P(x,y)=P_1(x_1,y)~|~P_2(x_2,y)
\ee
where $P_i(x_i,y):\mb{P},~i=1,2$, are the operands of $P(x,y)$.
If $P_i(x_i,y),~i=1,2$, do not contain disjunctions then $P(x,y)$ will be computable for the value assigned input lists $x_i,~i=1,2$, if at least one of $P_i(x_i,y),~i=1,2$, does not halt with a type violation error and returns an output.
Otherwise $P(x,y)$ will halt with a disjunction violation error.
Disjunctions that are computable in this sense are said to override type violation errors.
The input list of the main program $P(x,y)$ is defined in terms of the input lists ($x_i$, $i=1,2$) of its operand programs by adopting the convention
\be
x = x^\prime \setminus (x^\prime \bigcap y), \qquad x^\prime \simeq [x_1,x_2]
\ee 

\end{definition}
\noindent
The programs $P_i(x_i,y),~i=1,2$, may themselves represent program lists and $x^\prime \simeq [x_1,x_2]$ means that $x^\prime$ is the standard list concatenation of the lists $x_1,x_2$, with repeated variable names removed by following the procedure outlined in \cite{pan}.

\vspace{5mm}
\noindent
\textbf{Program equivalence.}
We now extend the definition of program equivalence to include disjunctions.

\begin{definition}(Program equivalence.)
Two programs $u:\mb{P}$ and $v:\mb{P}$ are said to be program equivalent provided that any of the following conditions is satisfied.
\begin{itemize}

\item $u=v$ ($u$ and $v$ are dummy variables representing the same program).

\item For some $n:\mb{I}_+$, $u=[u_i]_{i=1}^n$ and $v=[v_i]_{i=1}^n$ such that $[v_i]_{i=1}^n$ is a list permutation of $[u_i]_{i=1}^n$.   

\item $u=a~|~b$, $v=b~|~a$.

\item $u=[p,a|b]$, $v=[p,a]~|~[p,b]$.

\item $u=[a|b,p]$, $v=[a,p]~|~[b,p]$.

\item $v=[~]~|~u$.

\item $v=[[~],u]$.

\item $v=[u,[~]]$.

\end{itemize}
\noindent
Program equivalence is denoted by $u \equiv v$ and satisfies the properties of reflexivity, symmetry and transitivity.

\end{definition}

\vspace{5mm}
\noindent
\textbf{Operand programs.}
A program $s$ containing a disjunction can be expressed in the general form $s=[p,a_1 ~|~ \cdots ~|~ a_n,q]$, where $a_i:\mb{P},~i=1,\ldots,n$, and $p$ and/or $q$ may be the empty list program $[~]$.
We note that $a_1 ~|~ \cdots ~|~ a_n$ is not a program list but rather an element of the program list $[p,a_1 ~|~ \cdots ~|~ a_n,q]$. 
The program $s$ can be split into the programs $[p,a_i,q],~i=1,\ldots,n$, by the following two step procedure
\be
[p,a_1 ~|~ \cdots ~|~ a_n,q] \to \big [ [p,a_1] ~|~ \cdots ~|~ [p,a_n],q \big ] \to [p,a_1,q] ~|~ \cdots ~|~ [p,a_n,q]
\ee 
The programs $[p,a_i,q],~i=1,\ldots,n$, will be referred to as the operand programs of $s$ based on the disjunction $a_1 ~|~ \cdots ~|~ a_n$.
Sublist derivations can be performed independently on each operand program.
When independent derivations of the operand programs yield a common conclusion, say $c$, then that common conclusion can be contracted back onto the main program $s$ to produce a computable program extension $[s,c]$.
The disjunction contraction rule demonstrates how this is done.
There are two additional disjunction contraction rules that involve type $\mb{P}_{false}$ operand programs.
Note that the contraction rules involve two operands but can be generalized to include more than two operands. 
 
\vspace{5mm}
\noindent
\textbf{Disjunction contraction rules.}
To the existing construction rules we introduce the additional rule for programs containing disjunctions.

\vspace{5mm}
\noindent
Disjunction contraction rule.

\begin{crule}\label{dcr}
\be
\bal
& \Big [ \big [ Cpe([a,c],[~]),~Cpe([b,c],[~]),~Disj([a,b],[d]) \big ],~Acpe([d,c],[~]) \Big ] \\
\eal
\ee
\end{crule}

\noindent
The following contraction rules involve false programs.
They are not stated as higher order axioms because they can be derived as theorems.

\vspace{5mm}
\noindent
Disjunction contraction rule 2.

\be
\bal
& \Big [ \big [ False([a],[~]),~Cpe([b,c],[~]),~Disj([a,b],[d]) \big ],~Acpe([d,c],[~]) \Big ] \\
\eal
\ee

\vspace{5mm}
\noindent
Disjunction contraction rule 3.

\be
\bal
& \Big [ \big [ False([a],[~]),~False([b],[~]),~Disj([a,b],[d]) \big ],~Afalse([d],[~]) \Big ] \\
\eal
\ee

\vspace{5mm}
\noindent
\textbf{Execution errors.}
So far we have associated execution errors with type violations.
We now give an extended definition of an execution error that includes programs containing disjunctions.

\begin{definition}(Execution error.)
A program $s:\mb{P}$ will halt with an execution error if any of the following conditions occur.

\begin{itemize}

\item $s$ does not contain a disjunction and there is a type violation of any assigned value of the elements of its I/O lists.

\item $s$ contains a disjunction of the form $s=[p,a_1 ~|~ \cdots ~|~ a_n,q]$, where the programs $p$, $q$ and $a_i,~i=1,\ldots,n$, do not contain any disjunctions, and there is a type violation of an assigned value of the elements of the I/O lists of every operand program $[p,a_i,q],~i=1,\ldots,n$.
The program $s$ will not halt with an execution error if there are no type violations of any assigned value of the elements of the I/O lists in at least one operand program.
 
\end{itemize}

\end{definition}

\vspace{5mm}
\noindent
\textbf{Disjunction splitting.}
Let $p=[p_i]_{i=1}^m$, $q=[q_i]_{i=1}^n$ and consider the program $s=[p,d,q]$, where $d=a|b$ and $a=[a_i]_{i=1}^k$, $b=[b_i]_{i=1}^l$. 
Based on the disjunction $d$, the main program $[p,d,q]$ can be split into the two operand programs $u=[p,a,q]$ and $v=[p,b,q]$.
Suppose that we have independently applied sublist derivations to each of the parallel programs $u$ and $v$ to obtain a common conclusion $c$.
We may then contract the common conclusion $c$ back onto the main program by applying the disjunction contraction rule. 
The procedure is depicted in the following table. 

\be
\begin{array}{lllllllll}
label  & [s,c]   &                 & & label   & [u,c]   &               & label   & [v,c]   \\
       &         &                 & &         &         &               &         &         \\
1      & p_1     & \hspace{15mm}   & & 1       & p_1     & \hspace{10mm} & 1       & p_1     \\
\vdots & \vdots  & \hspace{15mm}   & & \vdots  & \vdots  & \hspace{10mm} & \vdots  & \vdots  \\
m      & p_m     & \hspace{15mm}   & & m       & p_m     & \hspace{10mm} & m       & p_m     \\
m+1    & d       & *               & & m+1     & a_1     & \hspace{10mm} & m+1     & b_1     \\
m+2    & q_1     & \hspace{15mm}   & & \vdots  & \vdots  & \hspace{10mm} & \vdots  & \vdots  \\
\vdots & \vdots  & \hspace{15mm}   & & m+k     & a_k     & \hspace{10mm} & m+l     & b_l     \\
m+n+1  & q_n     & \lra            & & m+k+1   & q_1     & \hspace{10mm} & m+l+1   & q_1     \\
m+n+2  & c       & \hspace{15mm}   & & \vdots  & \vdots  & \hspace{10mm} & \vdots  & \vdots  \\
       &         & \hspace{15mm}   & & m+k+n   & q_n     & \hspace{10mm} & m+l+n   & q_n     \\
       &         & & \lla          & m+k+n+1   & c       & \hspace{10mm} & m+l+n+1 & c       \\
\end{array}
\ee

The right arrow, $\lra$, indicates that the user has requested that the main program be split into two parallel operand programs at line $m+n+1$ of the main program based on the operands of the disjunction $d=a|b$.
The asterix next to the statement $d$ indicates that the disjunction splitting is based on the operands of that statement.
The left arrow, $\lla$, indicates that the common conclusion $c$ of the two parallel programs, $u$ and $v$, is to be contracted back onto the main program at line labeled $n+m+2$ of the main program by applying the disjunction contraction rule.

Suppose that under a sublist derivation the conclusion $c$ of $u$ and $v$, respectively, were derived from axioms/theorems labeled $A_1$ and $A_2$, respectively. 
Upon output the statement $c$ of the main program will have an attached composite connection list
\be
\big [ A_1,i_1,\ldots,i_{j_1} \big ] \big [ A_2,i_1^\prime,\ldots,i_{j_2}^\prime \big ]
\ee
where $1 \leq i_1,\ldots,i_{j_1} \leq m+k+n$ are labels of some sublist of the operand program $u$ and $1 \leq i_1^\prime,\ldots,i_{j_2}^\prime \leq m+l+n$ are labels of some sublist of the operand program $v$.
Here $j_1$ and $j_2$, respectively, are the lengths of the premise program lists of the axioms/theorems $A_1$ and $A_2$, respectively.

A contraction of the conclusion $c$ to the main program can also occur if one of the operand programs leads to a conclusion $c$ and the other a conclusion $False$.
This case follows from the disjunction contraction rule 2.
If both operand programs are type $\mb{P}_{false}$ then by the disjunction contraction rule 3 the main program will be assigned the type $\mb{P}_{false}$.
All of the above procedures can be extended in the obvious way where $d$ is a disjunction comprised of more than two operands.

In mainstream mathematics derivations associated with each operand are often conducted as separate cases within a single proof.
The reason for this is that many of these derivations are not of sufficient interest to be considered as separate theorems.
When using \emph{VPC}, derivations of proofs associated with each operand program must be conducted outside of the main proof program containing the disjunction.
The theorems extracted from the separate operand program derivations are to be stored in \emph{axiom.dat}. 
The derivation of the proof associated with the main program containing the disjunction can then access the theorems associated with each operand program through the disjunction contraction rules. 

Extracting theorems associated with each operand program may lead to an accumulation of theorems in \emph{axiom.dat} that are of a trivial nature and are not of particular interest in themselves.
However, this should not be a problem for storage and retrieval purposes.
In \emph{VPC} one may choose to label these theorems as lemmas to weaken their status.
There is sometimes an advantage in storing these individual operand cases as separate lemmas outside of the main proof because it is not uncommon that they are reused in later proofs.

\vspace{5mm}
\noindent
\textbf{Redundancy in disjunction splitting.}
Suppose that under disjunction splitting $v$ is a false program and $u$ is a computable program.
It is possible that we may find a sublist derivation leading to the conclusion of $v$ that coincides with the conclusion of $u$ before detecting the falsity of $v$.
We may then proceed to contract the common conclusion $c$ to the main proof program containing the disjunction to obtain $[[p,d,q],c]$.
We may suspect that this will lead to an error in our derivation of the main proof.
This will not be the case since, under the disjunction contraction rule 2, this would be the identical conclusion that would have been made if we had detected that $v:\mb{P}_{false}$.

\vspace{5mm}
\noindent
\textbf{Notes.}

\begin{itemize}

\item There remains the possibility that both operand programs are of type $\mb{P}_{false}$ and that derivations associated with both operand programs have been terminated prematurely with a common derived conclusion, say $c$.
We may then proceed to contract this common conclusion back onto the main program to obtain $[[p,d,q],c]$.
By the disjunction contraction rule 3 the derivations associated with each operand program should have been continued until one arrives at a common conclusion $False$ so that the program $[p,d,q]$ is identified as type $\mb{P}_{false}$.
This type of occurrence is related to the situation described in a note of the previous section and will be discussed further in a later section.

\end{itemize}

\section{Additional higher order construction axioms.}

Some of the following construction rules are in the form of existence axioms.
Others are given in two parts with an existence axiom followed by an equivalence relationship.
The disjunction distributivity rules are split into left and right, each involving two existence axioms followed by an equivalence axiom.
The empty list program is denoted by $ep=[~]$ and serves as a constant for type $\mb{P}$ objects.  

\vspace{5mm}
\noindent
\textbf{I/O type $\mb{P}$}.
An important property of all programs is that the type of the assigned values of all elements of the I/O lists are checked within the program.
The following axioms reflect this property for higher order programs. 

\begin{crule}\label{iot1}
\be
\big [ P(x,y),~Prog([p],[~]) \big ]  \qquad p \in x
\ee
\end{crule}

\begin{crule}\label{iot2}
\be
\big [ P(x,y),~Prog([p],[~]) \big ]  \qquad p \in y
\ee
\end{crule}
\noindent
Here $x$ and $y$ are lists and $P$ is a generic higher order program name.
The symbol $\in$ is used here to denote element of a list.

\vspace{5mm}
\noindent
\textbf{Axiom of substitution.}
To present the axiom of program substitution in a more general form we introduce the nonatomic list program $Eqvlst$.
For two lists $u=[u_i]_{i=1}^n$ and $v=[v_i]_{i=1}^n$, for some $n:\mb{I}_+$, we define the list
\be
\langle u,v \rangle = [u_1,v_1,u_2,v_2,\ldots,u_n,v_n]
\ee
The program $Eqvlst$ accepts input lists of the form $\langle u,v \rangle$ and represents a list of atomic program equivalence checking programs,
\be
Eqvlst( \langle u,v \rangle ,[~]) = \Big [ Equiv([u_i,v_i],[~]) \Big ]_{i=1}^n
\ee
For the case $n=1$ $Eqvlst$ reduces to the atomic program $Equiv$, i.e. $Eqvlst( \langle u,v \rangle ,[~]) = Equiv([u,v],[~])$.

The first part of the axiom of substitution is an existence axiom with some restrictions.
\begin{crule}\label{subst1}
\be
\Big [ \big [ P(x,y),~Eqvlst( \langle x^\prime,x \rangle,[~]) \big ],~P(x^\prime,y^\prime) \Big ] \hspace{10mm} P \notin [Cpe,False,Eqio] 
\ee
\end{crule}
\noindent
Here $x$, $x^\prime$, $y$ and $y^\prime$ are lists and $P$ is a generic higher order atomic program name excluding $Cpe,False,Eqio$.
Programs $Cpe$ and $False$ have their own special equivalence type assignment existence axioms (see CR2 and CR6).
Both $y$ and $y^\prime$ may be the empty list.
The second part of the axiom of substitution is applicable when $y$ and $y^\prime$ are not empty lists.
\begin{crule}\label{subst2}
\be
\Big [ \big [ P(x,y),~Eqvlst(\langle x^\prime,x \rangle,[~]),~P(x^\prime,y^\prime) \big ],~Eqvlst( \langle y^\prime,y \rangle ,[~]) \Big ]
\ee 
\end{crule}

\noindent
Here $P(x,y)$ may be any higher order atomic program with a nonempty output list.
For the case that $y$ and $y^\prime$ are lists of unit length we have $Eqvlst( \langle y^\prime,y \rangle ,[~]) = Equiv([y^\prime,y],[~])$. 

\vspace{5mm}
\noindent
\textbf{Program equivalence.}

\begin{crule}\label{equiv1}
\be
\bal
& \Big [ Prog([p],[~]),~Equiv([p,p],[~]) \Big ] \\
\eal
\ee
\end{crule}

\begin{crule}\label{equiv2}
\be
\bal
& \Big [ Equiv([p,q],[~]),~Equiv([q,p],[~]) \Big ] \\
\eal
\ee
\end{crule}

\noindent
Program equivalence satisfies the transitivity condition
\be
\bal
& \Big [ \big [ Equiv([p,q],[~]),~Equiv([q,r],[~]) \big ],~Equiv([p,r],[~]) \Big ] \\
\eal
\ee
This is not included as an axiom because it can be derived as a theorem using the axiom of substitution.

\vspace{5mm}
\noindent
\textbf{Program concatenation.}

\begin{crule}\label{pc1}
\be
\bal
& \Big [ Cpe([p,c],[~]),~Conc([p,c],[s]) \Big ] \\
\eal
\ee
\end{crule}

\noindent
Concatenation with the empty program.
\begin{crule}\label{pc2}
\be
\bal
& \Big [ Prog([p],[~]),~Conc([p,ep],[s]) \Big ] \\
\eal
\ee
\end{crule}
\begin{crule}\label{pc3}
\be
\bal
& \Big [ Conc([p,ep],[s]),~Equiv([s,p],[~]) \Big ] \\
\eal
\ee
\end{crule}
\begin{crule}\label{pc4}
\be
\bal
& \Big [ Prog([p],[~]),~Conc([ep,p],[s]) \Big ] \\
\eal
\ee
\end{crule}
\begin{crule}\label{pc5}
\be
\bal
& \Big [ Conc([ep,p],[s]),~Equiv([s,p],[~]) \Big ] \\
\eal
\ee
\end{crule}

\vspace{5mm}
\noindent
\textbf{Disjunctions.}

\vspace{5mm}
\noindent
Disjunction Commutativity.

\begin{crule}\label{dcomm1}
\be
\bal
& \Big [ Disj([a,b],[s]),~Disj([b,a],[r]) \Big ] \\
\eal
\ee
\end{crule}
\begin{crule}\label{dcomm2}
\be
\bal
& \Big [ \big [ Disj([a,b],[s]),~Disj([b,a],[r]) \big ], Equiv([r,s],[~]) \Big ] \\
\eal
\ee
\end{crule}

\vspace{5mm}
\noindent
Disjunction right distributivity.

\begin{crule}\label{drd1}
\be
\bal
& \Big [ \big [ Conc([p,a],[r]),~Conc([p,b],[s]),~Disj([a,b],[d]) \big ],~Disj([r,s],[v]) \Big ] \\
\eal
\ee
\end{crule}
\begin{crule}\label{drd2}
\be
\bal
& \Big [ \big [ Conc([p,a],[r]),~Conc([p,b],[s]),~Disj([a,b],[d]) \big ],~Conc([p,d],[u]) \Big ] \\
\eal
\ee
\end{crule}
\begin{crule}\label{drd3}
\be
\bal
& \Big [ \big [ Conc([p,a],[r]),~Conc([p,b],[s]),~Disj([a,b],[d]),~Conc([p,d],[u]),~Disj([r,s],[v]) \big ],~Equiv([u,v],[~]) \Big ] \\
\eal
\ee
\end{crule}

\vspace{5mm}
\noindent
Disjunction left distributivity.
\begin{crule}\label{dld1}
\be
\bal
& \Big [ \big [ Conc([a,p],[r]),~Conc([b,p],[s]),~Disj([a,b],[d]) \big ],~Disj([r,s],[v]) \Big ] \\
\eal
\ee
\end{crule}
\begin{crule}\label{dld2}
\be
\bal
& \Big [ \big [ Conc([a,p],[r]),~Conc([b,p],[s]),~Disj([a,b],[d]) \big ],~Conc([d,p],[u]) \Big ] \\
\eal
\ee
\end{crule}
\begin{crule}\label{dld3}
\be
\bal
& \Big [ \big [ Conc([a,p],[r]),~Conc([b,p],[s]),~Disj([a,b],[d]),~Conc([d,p],[u]),~Disj([r,s],[v]) \big ],~Equiv([u,v],[~]) \Big ] \\
\eal
\ee
\end{crule}

\vspace{5mm}
\noindent
Empty program operand.

\begin{crule}\label{epo1}
\be
\bal
& \Big [ Prog([p],[~]),~Disj([p,ep],[s]) \Big ] \\
\eal
\ee
\end{crule}
\begin{crule}\label{epo2}
\be
\bal
& \Big [ Disj([p,ep],[s]),~Equiv([s,p],[~]) \Big ] \\
\eal
\ee
\end{crule}

\vspace{5mm}
\noindent
False program operand.

\begin{crule}\label{fpo}
\be
\bal
& \Big [ \big [ Disj([a,b],[p]),~False([b],[~]) \big ], Equiv([p,a],[~]) \Big ] \\
\eal
\ee
\end{crule}

\vspace{5mm}
\noindent
\textbf{Notes.}

\begin{itemize}

\item The two separate rules, CR16 and CR18, for the existence of a concatenation with the empty program are necessary because we have no rule for commutativity of program concatenation.
On the other hand we do have a commutativity rule for disjunctions so that a single existence axiom for the disjunction with the empty program suffices.

\item The construction rules CR1-CR30 are presented as irreducible computable program extensions of higher order programs.
They can be regarded as the axioms of a theory for the construction of programs as proofs in the context of the formal system on which \emph{VPC} is based.
It would be useful to employ \emph{VPC} as a self referencing tool for the analysis and generalization of the construction rules themselves.
For reasons of brevity such a study will be presented elsewhere.
Instead we will proceed to the continuation of some basic results of arithmetic on $\mb{I}$ that was initiated in \cite{pan}.

\end{itemize}

\section{Arithmetic on $\mb{I}$}

As a demonstration of \emph{VPC} a number of basic identities and inequalities on $\mb{I}$ were derived in \cite{pan}.
Here we will demonstrate the additional features of \emph{VPC} that incorporate rules of falsity and disjunctions.
We use the same axioms of arithmetic on $\mb{I}$ as presented in \cite{pan} with some additions.
For arithmetic on $\mb{I}$ we employ the following atomic integer programs that are defined in the appendix.

\vspace{5mm}

\begin{tabular}{|c|c|}
\hline
Atomic program names & Atomic program type \\
\hline
$Int,~Lt,~Eq$ & $\mb{P}_{type}$ \\
\hline
$Aid,~Add,~Mult,~Div$ & $\mb{P}_{assign}$ \\
\hline
\end{tabular} 

\vspace{5mm}
\noindent
The program $Neq$ has been removed from the collection of atomic programs as presented in \cite{pan} and will be redefined later as a disjunction.

Axioms for arithmetic on $\mb{I}$ are labeled with an upper case A followed by a number.
These axioms are stored in a file $axiom.dat$ that is accessed by \emph{VPC} during proof construction.
In the following axioms the internal square brackets delimit the premise program from the conclusion.
(The standard list concatenation for programs apply so that the internal brackets can be removed.)
When there is only a single premise the internal square brackets are omitted.

\vspace{5mm}
\noindent
\textbf{I/O type $\mb{I}$.}
An important property of all programs is that the type of the assigned values of all elements of the I/O lists are checked within the program.
The following axioms reflect this property for integer programs.

\begin{axm}\label{ioti1}
\be
\big [ P(x,y),~Int([c],[~]) \big ]  \qquad c \in x
\ee
\end{axm}
\begin{axm}\label{ioti2}
\be
\big [ P(x,y),~Int([c],[~]) \big ]  \qquad c \in y
\ee
\end{axm}

\noindent
Here $x$ and $y$ are lists and $P$ is a generic integer program name. 

\vspace{5mm}
\noindent
\textbf{Axiom of substitution for integer programs.}
The axiom of substitution for integer programs is almost identical to that of higher order programs.
The program equivalence checking programs are replaced by the integer equality checking programs.
For two lists $u=[u_i]_{i=1}^n$ and $v=[v_i]_{i=1}^n$, for some $n:\mb{I}_+$, we define the list
\be
\langle u,v \rangle = [u_1,v_1,u_2,v_2,\ldots,u_n,v_n]
\ee
The program $Eqlst$ accepts input lists of the form $\langle u,v \rangle$ and represents a list of atomic equality checking programs,
\be
Eqlst( \langle u,v \rangle ,[~]) = \Big [ Eq([u_i,v_i],[~]) \Big ]_{i=1}^n
\ee
For the case $n=1$ the list equality program $Eqlst$ reduces to the atomic program $Eq$, i.e. $Eqlst( \langle u,v \rangle ,[~]) = Eq([u,v],[~])$.
The first part of the axiom of substitution is an existence axiom.
\begin{axm}\label{substi1}
\be
\Big [ \big [ P(x,y),~Eqlst( \langle x^\prime,x \rangle,[~]) \big ],~P(x^\prime,y^\prime) \Big ]
\ee
\end{axm}
\noindent
Here $x$, $x^\prime$, $y$ and $y^\prime$ are lists and $P$ is a generic integer program name.
Both $y$ and $y^\prime$ may be the empty list.
The second part of the axiom of substitution is applicable when $y$ and $y^\prime$ are not empty lists.
\begin{axm}\label{substi2}
\be
\Big [ \big [ P(x,y),~Eqlst( \langle x^\prime,x \rangle,[~]),~P(x^\prime,y^\prime) \big ],~Eqlst( \langle y^\prime,y \rangle ,[~]) \Big ]
\ee 
\end{axm}
\noindent
For the case that $y$ and $y^\prime$ are lists of unit length we have $Eqlst( \langle y^\prime,y \rangle ,[~]) = Eq([y^\prime,y],[~])$. 

\vspace{5mm}
\noindent
\textbf{Equality axioms.}

\noindent
Reflexivity.
\begin{axm}\label{eql1}
\be
\big [ Int([a],[~]),~Eq([a,a],[~]) \big ]
\ee
\end{axm}

\noindent
Symmetry.
\begin{axm}\label{eql2}
\be
\big [ Eq([a,b],[~]),~Eq([b,a],[~]) \big ]
\ee
\end{axm}

\noindent
Transitivity of equality states that
\be
\Big [ \big [ Eq([a,b],[~]),~Eq([b,c],[~]) \big ],~Eq([a,c],[~]) \Big ]
\ee
We do not include this as an axiom because it can be derived from the first part of the axiom of substitution.

\vspace{5mm}
\noindent
\textbf{Identity assignment axiom.}

\begin{axm}\label{as}
\be
\big [ Aid([a],[b]),~Eq([b,a],[~]) \big ]
\ee
\end{axm}

\vspace{5mm}
\noindent
\textbf{Notes.}

\begin{itemize}

\item Here we have slightly modified the axiom of substitution of \cite{pan}.
Both versions of the axiom of substitution can be derived from each other and proofs using either of these axioms are similar.

\item Axiom A4 is an example where the conclusion is a nonatomic program list.
For our purposes we will only require the case where the lengths of the output lists $y$ and $y^\prime$ have unit length so that the conclusion reduces to the atomic program $Eq([y^\prime,y],[~])$.   

\end{itemize}

\vspace{5mm}
\noindent
\section{Axioms of arithmetic on $\mb{I}$.}

\vspace{5mm}
\noindent
\textbf{Axioms of addition and multiplication.}

\noindent
Commutativity of addition.

\begin{axm}\label{addcom1}
\be
\big [ Add([a,b],[c]),~Add([b,a],[d]) \big ]
\ee
\end{axm}

\begin{axm}\label{addcom2}
\be
\Big [ \big [ Add([a,b],[c]),~Add([b,a],[d]) \big ],~Eq([d,c],[~]) \Big ]
\ee
\end{axm}

\noindent
Associativity of addition.

\begin{axm}\label{addassoc1}
\be
\Big [ \big [ Add([a,b],[d]),~Add([d,c],[x]),~Add([b,c],[e]) \big ],~Add([a,e],[y]) \Big ]
\ee
\end{axm}

\begin{axm}\label{addassoc2}
\be
\bal
& \Big [ \big [ Add([a,b],[d]),~Add([d,c],[x]),~Add([b,c],[e]),~Add([a,e],[y]) \big ], Eq([y,x],[~]) \Big ] \\
\eal
\ee
\end{axm}

\noindent
Addition by zero.

\begin{axm}\label{addz1}
\be
\big [ Int([a],[~]),~Add([a,0],[b]) \big ]
\ee
\end{axm}
\begin{axm}\label{addz2}
\be
\big [ Add([a,0],[b]),~Eq([b,a],[~]) \big ]
\ee
\end{axm}

\noindent
Additive inverse.

\begin{axm}\label{multinv1}
\be
\big [ Int([a],[~]),~Mult([-1,a],[b]) \big ]
\ee
\end{axm}
\begin{axm}\label{multinv2}
\be
\big [ Mult([-1,a],[b]),~Add([a,b],[d]) \big ]
\ee
\end{axm}
\begin{axm}\label{multinv3}
\be
\Big [ \big [ Mult([-1,a],[b]),~Add([a,b],[d]) \big ],~Eq([d,0],[~]) \Big ]
\ee
\end{axm}

\noindent
Commutativity of multiplication.

\begin{axm}\label{multcomm1}
\be
\big [ Mult([a,b],[c]),~Mult([b,a],[d]) \big ]
\ee
\end{axm}
\begin{axm}\label{multcomm2}
\be
\Big [ \big [ Mult([a,b],[c]),~Mult([b,a],[d]) \big ],~Eq([d,c],[~]) \Big ]
\ee
\end{axm}

\noindent
Associativity of multiplication.

\begin{axm}\label{multassoc1}
\be
\Big [ \big [ Mult([a,b],[d]),~Mult([d,c],[x]),~Mult([b,c],e] \big ],~Mult([a,e],[y]) \Big ]
\ee
\end{axm}
\begin{axm}\label{multassoc2}
\be
\Big [ \big [ Mult([a,b],[d]),~Mult([d,c],[x]),~Mult([b,c],e],~Mult([a,e],[y]) \big ],~Eq([y,x],[~]) \Big ]
\ee
\end{axm}

\noindent
Multiplication by unity.

\begin{axm}\label{multun1}
\be
\big [ Int([a],[~]),~Mult([1,a],[b]) \big ]
\ee
\end{axm}
\begin{axm}\label{multun2}
\be
\big [ Mult([1,a],[b]),~Eq([b,a],[~]) \big ]
\ee
\end{axm}

\noindent
Distributive law.

\begin{axm}\label{dist1}
\be
\Big [ \big [ Add([b,c],[d]),~Mult([a,d],[x]),~Mult([a,b],[u]),~Mult([a,c],[v]) \big ],~Add([u,v],[y]) \Big ]
\ee
\end{axm}
\begin{axm}\label{dist2}
\be
\Big [ \big [ Mult([a,b],[u]),~Mult([a,c],[v]),~Add([u,v],[y]),~Add([b,c],[d]) \big ],~Mult([a,d],[x]) \Big ]
\ee
\end{axm}
\begin{axm}\label{dist3}
\be
\bal
& \Big [ \big [ Add([b,c],[d]),~Mult([a,d],[x]),~Mult([a,b],[u]),~Mult([a,c],[v]),~Add([u,v],[y]) \big ],~Eq([y,x],[~]) \Big ] \\
\eal
\ee
\end{axm}

\vspace{5mm}
\noindent
\textbf{Order axioms.}

\begin{axm}\label{ordadd}
\be
\Big [ \big [ Lt([a,b],[~]),~Add([a,c],[x]),~Add([b,c],[y]) \big ],~Lt([x,y],[~]) \Big ]
\ee
\end{axm}
\begin{axm}\label{ordmult1}
\be
\Big [ \big [ Lt([a,b],[~]),~Lt([0,c],[~]),~Mult([a,c],[x]),~Mult([b,c],[y]) \big ],~Lt([x,y],[~]) \Big ]
\ee
\end{axm}
\begin{axm}\label{ordmult2}
\be
\Big [ \big [ Lt([a,b],[~]),~Lt([c,0],[~]),~Mult([a,c],[x]),~Mult([b,c],[y]) \big ],~Lt([y,x],[~]) \Big ]
\ee
\end{axm}

\noindent
Transitivity.

\begin{axm}\label{ordtrans}
\be
\Big [ \big [ Lt([a,b],[~]),~Lt([b,c],[~]) \big ],~Lt([a,c],[~]) \Big ]
\ee
\end{axm}

\vspace{5mm}
\noindent
\textbf{Divisor.}

\begin{axm}\label{div1}
\be
\Big [ \big [ Neq([a,0],[~]),~Mult([a,b],[c]) \big ],~Div([c,a],[d]) \Big ]
\ee
\end{axm}
\begin{axm}\label{div2}
\be
\Big [ \big [ Mult([a,b],[c]),~Div([c,a],[d]) \big ],~Eq([d,b],[~]) \Big ]
\ee
\end{axm}

\vspace{5mm}
\noindent
\textbf{Additional order axiom.}
To the order axioms we include an axiom that has an empty list premise 
\begin{axm}\label{ordunz}
\be
Lt([0,1],[~])
\ee
\end{axm}

\vspace{5mm}
\noindent
\textbf{Axiom of falsity (higher order type checking axiom).}
Axioms of falsity are higher order constructs.
We depart from the convention slightly of expressing the axiom in the form of a concatenation of a premise and conclusion $[p,c]$ by simply assigning the premise as a type $p:\mb{P}_{false}$.
For arithmetic on $\mb{I}$ we include the following axiom of falsity.
\begin{axm}\label{falsity}
\be
Lt([a,a],[~])~:~\mb{P}_{false}
\ee
\end{axm}

\noindent
One can think of this as being equivalent to the higher order axiom with an empty premise list
\be
False([p],[~])
\ee
where the assigned value of $p$ is an object of type $\mb{P}$ and is given by $p=Lt([a,a],[~])$.

\vspace{5mm}
\noindent
\textbf{Notes.}

\begin{itemize}

\item Axioms A1-A33 are the same as in \cite{pan} except for the slight modification of the axiom of substitution and the addition of the two axioms A32 and A33.

\item We have departed slightly from the convention of representing all elements of program I/O lists with alphanumeric variable names by allowing some elements to be represented by the constants $-1,0,1$.
To strictly adhere to the convention we could introduce special alphanumeric names for these constants.
For convenience we allow these constants to appear in the input lists as exceptions. 

\item There is no concise way to construct for each program a corresponding program that can serve as its negation. 
This is because programs often contain several actions that include type checking of input variables as well as possible assignments.
For example the program $Add([a,b],[c])$ is equivalent to the statement that if $a:\mb{I}$, $b:\mb{I}$ then construct the assignment $c:=a+b$ provided that $a+b:\mb{I}$.
Otherwise halt with an execution error.
There is no straight forward way to define a program that would serve as a negation to $Add([a,b],[c])$ as a formal statement. 

\item I/O equivalence does not satisfy a general rule of symmetry.
For example the program $Add([a,a],[c])$ is I/O equivalent to the program $Add([a,b],[d])$ but $Add([a,b],[d])$ is not I/O equivalent to $Add([a,a],[c])$.

\end{itemize}

\section{Special nonatomic programs.}

\vspace{5mm}
\noindent
\textbf{Not equal.}
We redefine the program $Neq$ from an atomic program in \cite{pan} to the disjunction
\be
Neq([a,b],[~]) = Lt([a,b],[~])~|~Lt([b,a],[~])
\ee
This does not change any of the derivations presented in \cite{pan}.

\vspace{5mm}
\noindent
\textbf{Less than or equal.}
\be
Le([a,b],[~]) = Lt([a,b],[~])~|~Eq([a,b],[~])
\ee
We include the following generalization of the order axiom A26 (see notes below). 
\begin{axm}\label{ordadd2}
\be
\Big [ \big [ Le([a,b],[~]),~Le([c,d],[~]),~Add([a,c],[x]),~Add([b,d],[y]) \big ],~Le([x,y],[~]) \Big ]
\ee
\end{axm}

\vspace{5mm}
\noindent
\textbf{Absolute value.}

\be
Abs([x,0,-1],[y]) = \big [ Lt([x,0],[~]),~Mult([-1,x],[y]) \big ]~|~\big [ Le([0,x],[~]),~Aid([x],[y]) \big ]
\ee

\noindent
The program $Abs([x,0,-1],[y])$ makes the assignment $y:=|x|$.
The constants 0 and -1 appear in the input list of $Abs$ only to follow the convention used to define the input list of the main program in the formal definition of program disjunctions.
$Abs$ has no useful meaning here if these constants are replaced by arbitrary variable names.

\vspace{5mm}
\noindent
\textbf{Special nonatomic program axioms.}
We include the following axioms.
\begin{axm}\label{dwneq}
\be
\big [ Lt([a,b],[~]),~Neq([a,b],[~]) \big ]
\ee
\end{axm}
\begin{axm}\label{dwle1}
\be
\big [ Eq([a,b],[~]),~Le([a,b],[~]) \big ]
\ee
\end{axm}
\begin{axm}\label{dwle2}
\be
\big [ Lt([a,b],[~]),~Le([a,b],[~]) \big ]
\ee
\end{axm}

\vspace{5mm}
\noindent
\textbf{Notes.}

\begin{itemize}

\item One may regard A34 to be derivable from the statement that if $a<b$ and $c<d$ and $x=a+c$ and $y=b+d$ then $x<y$.
It should be noted that one cannot derive A26 from the latter statement nor does latter statement follow trivially from A26.
The reason for this is that it would require the inclusion of an intermediate sum $z=a+d$, the existence of which cannot be assumed on $\mb{I}$.

\item Axioms A35-A37 have some similarity with a more general rule of weakening by disjunction introduction of propositional logic.
In our formal system there is no straight forward way to define a higher order construction rule for disjunction introduction because a program cannot be weakened by a disjunction with just any other program.
This is largely due to the fact that programs are defined in terms of the variable names of their I/O lists and therefore are constrained by certain rules of composition.
For example the program $Add([a,b],[c])$ cannot be weakened by a disjunction with the program $Add([a,c],[d])$ because the output $c$ of the former would not be available to the latter.

\item Program lists may contain repeated subprograms only when those subprograms have an empty list output.
Because of efficiency of execution it is generally better to avoid including repetitions.
However, there are situations where this is not possible.
For example the program $[Lt([a,b],[~]),~Le([a,b],[~])]$ has an implied repetition under the equivalent representation $[Lt([a,b],[~]),~Lt([a,b],[~])]~|~[Lt([a,b],[~]),~Eq([a,b],[~])]$ obtained through the disjunction distributivity rules.
The first operand contains a repeated program but is computable.
The second operand is a false program.

\end{itemize}

\section{Basic inequalities on $\mb{I}$}

The aim of this section is to demonstrate the additional features of \emph{VPC} that include disjunctions and detecting programs of type $\mb{P}_{false}$.
The derivations below are to be regarded as a continuation of those presented in \cite{pan}.

With the use of disjunction splitting the proofs associated with the separate operand programs precede the proof of the main program containing the disjunction.
They correspond to the separate cases that are accessed by the proof of the main program containing the disjunction and are stored as lemmas.
Lemmas are labeled by an upper case L followed by a number.
A statement containing two connection lists indicates that disjunction splitting has been applied at the preceding line of the proof and that the statement itself is obtained from a contraction of the conclusions of the operand programs.
A statement followed by an asterix $*$ indicates that disjunction splitting has been applied to the operands of that statement.
It is important to note that, as illustrated in the table of the section on disjunctions, the subprogram labels in multiple connection lists correspond to the labels of the subprograms of each operand program that results from the disjunction splitting and not the labels associated with the main program.
These parallel programs along with their labels are not shown here due to space limitations although they are available as output in \emph{VPC}.     

\vspace{5mm}
\noindent
\textbf{Preliminaries.}
We start by presenting a few preliminary results that will be needed for later proofs.
They are largely based on those presented in \cite{pan} with a few additions.
The following derives the transitivity of equality by using the axiom of substitution.
\begin{lstlisting}
Theorem T1.
[ [ Eq([a,b],[]), Eq([b,c],[]) ], Eq([a,c],[]) ]

Proof.
  1 Eq([a,b],[])        
  2 Eq([b,c],[])        
  3 Int([a],[])         [A1,1]
  4 Eq([a,a],[])        [A5,3]
  5 Eq([a,c],[])        [A3,1,4,2]
\end{lstlisting}
The proofs of theorems of \cite{pan} were regenerated using the current version of the axiom of substitution.
The proofs are very similar to those given in \cite{pan} and the theorems are presented here without their proofs.
Note that with the inclusion of theorem T1 above the theorem labels T1-T15 of \cite{pan} are now relabelled to T2-T16.
\begin{lstlisting}
Theorem T2.
[ [ Add([a,b],[c]), Mult([-1,b],[d]) ], Add([c,d],[m]) ]

Theorem T3.
[ [ Add([a,b],[c]), Mult([-1,b],[d]), Add([c,d],[m]) ], Eq([m,a],[]) ]

Theorem T4.
[ [ Add([a,b],[c]), Add([a,d],[e]), Eq([c,e],[]) ], Eq([b,d],[]) ]

Theorem T5.
[ [ Mult([a,b],[c]), Mult([a,d],[e]), Eq([c,e],[]), Neq([a,0],[]) ],
 Eq([b,d],[]) ]

Theorem T6.
[ [ Int([a],[]) ], Mult([0,a],[n]) ]

Theorem T7.
[ [ Mult([0,a],[b]) ], Eq([b,0],[]) ]

Theorem T8.
[ [ Mult([-1,a],[b]), Mult([-1,b],[c]) ], Eq([c,a],[]) ]

Theorem T9.
[ [ Mult([a,b],[c]), Mult([-1,b],[d]) ], Mult([a,d],[i]) ]

Theorem T10.
[ [ Mult([a,b],[c]), Mult([-1,b],[d]), Mult([a,d],[i]), Mult([-1,c],[e]) ],
 Eq([i,e],[]) ]

Theorem T11.
[ [ Mult([a,b],[c]), Mult([-1,a],[d]) ], Mult([d,b],[g]) ]

Theorem T12.
[ [ Mult([a,b],[c]), Mult([-1,a],[d]), Mult([d,b],[g]), Mult([-1,c],[h]) ],
 Eq([g,h],[]) ]

Theorem T13.
[ [ Mult([a,b],[c]), Mult([-1,a],[d]), Mult([-1,b],[e]) ], Mult([d,e],[g]) ]

Theorem T14.
[ [ Mult([a,b],[c]), Mult([-1,a],[d]), Mult([-1,b],[e]), Mult([d,e],[f]) ],
 Eq([f,c],[]) ]

Theorem T15.
[ [ Lt([0,a],[]), Mult([-1,a],[b]) ], Lt([b,0],[]) ]

Theorem T16.
[ [ Lt([a,0],[]), Mult([-1,a],[b]) ], Lt([0,b],[]) ]
\end{lstlisting}
To the above theorems we include the following result that will also be needed in later derivations.
Like axiom A32, theorem T17 has an empty list premise.
\begin{lstlisting}
Theorem T17.
[ Lt([-1,0],[]) ]

Proof.
  1 Lt([0,1],[])        [A32]
  2 Int([1],[])         [A1,1]
  3 Mult([-1,1],[a])    [A14,2]
  4 Mult([1,-1],[b])    [A17,3]
  5 Eq([b,-1],[])       [A22,4]
  6 Eq([a,b],[])        [A18,4,3]
  7 Eq([a,-1],[])       [T1,6,5]
  8 Lt([a,0],[])        [T15,1,3]
  9 Int([0],[])         [A1,1]
 10 Eq([0,0],[])        [A5,9]
 11 Lt([-1,0],[])       [A3,8,7,10]
\end{lstlisting}

\vspace{5mm}
\noindent
\textbf{Disjunction contraction rule.}
As a first application of the contraction rule we relabel theorems T16-T17 of \cite{pan} to lemmas L1-L2.
\begin{lstlisting}
Lemma L1.
[ [ Lt([0,a],[]), Mult([a,a],[b]) ], Lt([0,b],[]) ]

Lemma L2.
[ [ Lt([a,0],[]), Mult([a,a],[b]) ], Lt([0,b],[]) ]
\end{lstlisting}
We now apply the contraction rule to obtain
\begin{lstlisting}
Theorem T18.
[ [ Neq([a,0],[]), Mult([a,a],[b]) ], Lt([0,b],[]) ]

Proof.
  1 Neq([a,0],[])*      
  2 Mult([a,a],[b])     
  3 Lt([0,b],[])        [L2,1,2][L1,1,2]
\end{lstlisting}
\emph{VPC} splits the above premise into the two operand programs $[Lt([0,a],[~]),~Mult([a,a],[b])]$ and $[Lt([a,0],[~]),~Mult([a,a],[b])]$.
A search is conducted for premises of the axioms/theorems stored in the file \emph{axiom.dat} that can be matched to sublists of each operand program and their conclusions stored in memory.
It then searches through the two collections of conclusions associated with each operand program and extracts those conclusions that are common to both.

Theorem T19 generalizes the order axiom of transitivity A29 and can be translated to the statement that if $a \leq b$ and $b \leq c$ then $a \leq c$.
Theorem T19 is preceded by two lemmas, L4 and L5, that are associated with the separate cases for the disjunction $Le$ contained in T19.
The common conclusion of the two lemmas is contracted back onto the main proof of T19.
Lemma L4 involves a disjunction splitting and a contraction of the common conclusions obtained by employing axiom A29 and L3.  
\begin{lstlisting}
Lemma L3.
[ [ Lt([a,b],[]), Eq([b,c],[]) ], Lt([a,c],[]) ]

Proof.
  1 Lt([a,b],[])        
  2 Eq([b,c],[])        
  3 Int([a],[])         [A1,1]
  4 Eq([a,a],[])        [A5,3]
  5 Lt([a,c],[])        [A3,1,4,2]

Lemma L4.
[ [ Lt([a,b],[]), Le([b,c],[]) ], Le([a,c],[]) ]

Proof.
  1 Lt([a,b],[])        
  2 Le([b,c],[])*       
  3 Lt([a,c],[])        [A29,1,2][L3,1,2]
  4 Le([a,c],[])        [A37,3]

Lemma L5.
[ [ Eq([a,b],[]), Le([b,c],[]) ], Le([a,c],[]) ]

Proof.
  1 Eq([a,b],[])        
  2 Le([b,c],[])        
  3 Eq([b,a],[])        [A6,1]
  4 Int([c],[])         [A1,2]
  5 Eq([c,c],[])        [A5,4]
  6 Le([a,c],[])        [A3,2,3,5]

Theorem T19.
[ [ Le([a,b],[]), Le([b,c],[]) ], Le([a,c],[]) ]

Proof.
  1 Le([a,b],[])*       
  2 Le([b,c],[])        
  3 Le([a,c],[])        [L4,1,2][L5,1,2]
\end{lstlisting}
Theorem T20 generalizes theorem T15 and can be translated to the statement that if $c \geq 0$ then $-c \leq 0$.
The separate cases associated with the operand programs resulting from the disjunction splitting of $Le$ are given by the preceding lemmas L6 and L7.
The common conclusion of the two lemmas is contracted back onto the main proof of T20.
\begin{lstlisting}
Lemma L6.
[ [ Lt([0,c],[]), Mult([-1,c],[d]) ], Le([d,0],[]) ]

Proof.
  1 Lt([0,c],[])        
  2 Mult([-1,c],[d])    
  3 Lt([d,0],[])        [T15,1,2]
  4 Le([d,0],[])        [A37,3]

Lemma L7.
[ [ Eq([0,c],[]), Mult([-1,c],[d]) ], Le([d,0],[]) ]

Proof.
  1 Eq([0,c],[])        
  2 Mult([-1,c],[d])    
  3 Eq([c,0],[])        [A6,1]
  4 Int([-1],[])        [A1,2]
  5 Eq([-1,-1],[])      [A5,4]
  6 Mult([-1,0],[a])    [A3,2,5,3]
  7 Mult([0,-1],[b])    [A17,6]
  8 Eq([b,a],[])        [A18,6,7]
  9 Eq([b,0],[])        [T7,7]
 10 Int([0],[])         [A1,1]
 11 Eq([0,0],[])        [A5,10]
 12 Eq([a,0],[])        [A3,9,8,11]
 13 Eq([d,a],[])        [A4,2,5,3,6]
 14 Eq([d,0],[])        [T1,13,12]
 15 Le([d,0],[])        [A36,14]

Theorem T20.
[ [ Le([0,c],[]), Mult([-1,c],[d]) ], Le([d,0],[]) ]

Proof.
  1 Le([0,c],[])*       
  2 Mult([-1,c],[d])    
  3 Le([d,0],[])        [L6,1,2][L7,1,2]
\end{lstlisting}

\vspace{5mm}
\noindent
\textbf{Eliminating false programs in disjunctions.}
Here we present examples where a disjunction splitting and contraction involves detecting a false program in one of the operand program proofs.
Theorem T21 is equivalent to the statement that if $|a|=0$ then $a=0$.
It is preceded by two lemmas, L8 and L9, that are associated with the two operand programs that result from disjunction splitting in T21.
The premise of the first lemma, L8, is type $\mb{P}_{false}$.
Appealing to the disjunction contraction rule 2, the conclusion of the second lemma, L9, is contracted back onto the main proof of T21.
\begin{lstlisting}
Lemma L8.
[ Lt([a,0],[]) Mult([-1,a],[b]), Eq([b,0],[]) ]:False

Proof.
  1 Lt([a,0],[])        
  2 Mult([-1,a],[b])    
  3 Eq([b,0],[])        
  4 Lt([0,b],[])        [T16,1,2]
  5 Lt([0,0],[])        [L3,4,3]
  6 False               [A33,5]

Lemma L9.
[ [ Aid([a],[b]), Eq([b,0],[]) ], Eq([a,0],[]) ]

Proof.
  1 Aid([a],[b])        
  2 Eq([b,0],[])        
  3 Eq([b,a],[])        [A7,1]
  4 Int([0],[])         [A1,2]
  5 Eq([0,0],[])        [A5,4]
  6 Eq([a,0],[])        [A3,2,3,5]

Theorem T21.
[ [ Abs([a,0,-1],[b]), Eq([b,0],[]) ], Eq([a,0],[]) ]

Proof.
  1 Abs([a,0,-1],[b])*  
  2 Eq([b,0],[])        
  3 Eq([a,0],[])        [L8,1,2,3][L9,2,3]
\end{lstlisting}
Theorem T22 is the converse of T21 and is equivalent to the statement that if $a=0$ then $|a|=0$.
It is preceded by two lemmas, L10 and L11, that are associated with the two operand programs that result from disjunction splitting in T22.
The premise of one of the lemmas, L10, is type $\mb{P}_{false}$ so that the conclusion of L11 is contracted back onto the main proof of T22. 
\begin{lstlisting}
Lemma L10.
[ Lt([a,b],[]) Eq([a,b],[]) ]:False

Proof.
  1 Lt([a,b],[])        
  2 Eq([a,b],[])        
  3 Int([b],[])         [A1,1]
  4 Eq([b,b],[])        [A5,3]
  5 Lt([b,b],[])        [A3,1,2,4]
  6 False               [A33,5]

Lemma L11.
[ [ Aid([a],[b]), Eq([a,0],[]) ], Eq([b,0],[]) ]

Proof.
  1 Aid([a],[b])        
  2 Eq([a,0],[])        
  3 Eq([b,a],[])        [A7,1]
  4 Eq([b,0],[])        [T1,3,2]

Theorem T22.
[ [ Abs([a,0,-1],[b]), Eq([a,0],[]) ], Eq([b,0],[]) ]

Proof.
  1 Abs([a,0,-1],[b])*  
  2 Eq([a,0],[])        
  3 Eq([b,0],[])        [L10,1,3][L11,2,3]
\end{lstlisting}

\vspace{5mm}
\noindent
\textbf{Triangle inequality.}
For reasons of brevity we will skip some of the standard inequalities for absolute values and go straight to the proof of the triangle inequality.
We start by proving that if $|a| \leq c$ then $-c \leq a \leq c$.
Theorems T23 and T24, respectively, split this into the two parts leading to the conclusions $a \leq c$ and $-c \leq a$, respectively.
Theorem T23 is preceded by the two lemmas L12 and L13 that are associated with the two operand programs that result from disjunction splitting in T23.
\begin{lstlisting}
Lemma L12.
[ [ Lt([a,0],[]), Mult([-1,a],[b]), Le([b,c],[]) ], Le([a,c],[]) ]

Proof.
  1 Lt([a,0],[])        
  2 Mult([-1,a],[b])    
  3 Le([b,c],[])        
  4 Lt([0,b],[])        [T16,1,2]
  5 Lt([a,b],[])        [A29,1,4]
  6 Le([a,c],[])        [L4,5,3]

Lemma L13.
[ [ Aid([a],[b]), Le([b,c],[]) ], Le([a,c],[]) ]

Proof.
  1 Aid([a],[b])        
  2 Le([b,c],[])        
  3 Eq([b,a],[])        [A7,1]
  4 Int([c],[])         [A1,2]
  5 Eq([c,c],[])        [A5,4]
  6 Le([a,c],[])        [A3,2,3,5]

Theorem T23.
[ [ Abs([a,0,-1],[b]), Le([b,c],[]) ], Le([a,c],[]) ]

Proof.
  1 Abs([a,0,-1],[b])*  
  2 Le([b,c],[])        
  3 Le([a,c],[])        [L12,1,2,3][L13,2,3]
\end{lstlisting}
Lemmas L16 and L17 are associated with the two operand programs that result from the disjunction splitting of $Abs([a,0,-1],[b])$ that appears in the premise of T24.
Lemma L16 depends on the two preceding lemmas L14 and L15 that are associated with the two operand programs that result from disjunction splitting of $Le([b,c],[~])$ that appears in the premise of L16.
\begin{lstlisting}
Lemma L14.
[ [ Mult([-1,a],[b]), Lt([b,c],[]), Mult([-1,c],[d]) ], Le([d,a],[]) ]

Proof.
  1 Mult([-1,a],[b])    
  2 Lt([b,c],[])        
  3 Mult([-1,c],[d])    
  4 Mult([-1,b],[e])    [T9,1,1]
  5 Eq([e,a],[])        [T8,1,4]
  6 Mult([b,-1],[f])    [A17,4]
  7 Mult([c,-1],[g])    [A17,3]
  8 Lt([-1,0],[])       [T17]
  9 Lt([g,f],[])        [A28,2,8,6,7]
 10 Eq([g,d],[])        [A18,3,7]
 11 Int([f],[])         [A2,6]
 12 Eq([f,f],[])        [A5,11]
 13 Eq([f,e],[])        [A18,4,6]
 14 Eq([f,a],[])        [A3,13,12,5]
 15 Lt([d,a],[])        [A3,9,10,14]
 16 Le([d,a],[])        [A37,15]

Lemma L15.
[ [ Mult([-1,a],[b]), Eq([b,c],[]), Mult([-1,c],[d]) ], Le([d,a],[]) ]

Proof.
  1 Mult([-1,a],[b])    
  2 Eq([b,c],[])        
  3 Mult([-1,c],[d])    
  4 Mult([-1,b],[e])    [T9,1,1]
  5 Eq([e,a],[])        [T8,1,4]
  6 Int([-1],[])        [A1,1]
  7 Eq([-1,-1],[])      [A5,6]
  8 Eq([e,d],[])        [A4,4,7,2,3]
  9 Eq([d,e],[])        [A6,8]
 10 Eq([d,a],[])        [T1,9,5]
 11 Le([d,a],[])        [A36,10]

Lemma L16.
[ [ Mult([-1,a],[b]), Le([b,c],[]), Mult([-1,c],[d]) ], Le([d,a],[]) ]

Proof.
  1 Mult([-1,a],[b])    
  2 Le([b,c],[])*       
  3 Mult([-1,c],[d])    
  4 Le([d,a],[])        [L14,1,2,3][L15,1,2,3]

Lemma L17.
[ [ Le([0,a],[]), Aid([a],[b]), Le([b,c],[]), Mult([-1,c],[d]) ],
 Le([d,a],[]) ]

Proof.
  1 Le([0,a],[])        
  2 Aid([a],[b])        
  3 Le([b,c],[])        
  4 Mult([-1,c],[d])    
  5 Le([a,c],[])        [L13,2,3]
  6 Le([0,c],[])        [T19,1,5]
  7 Le([d,0],[])        [T20,6,4]
  8 Le([d,a],[])        [T19,7,1]

Theorem T24.
[ [ Abs([a,0,-1],[b]), Le([b,c],[]), Mult([-1,c],[d]) ], Le([d,a],[]) ]

Proof.
  1 Abs([a,0,-1],[b])*  
  2 Le([b,c],[])        
  3 Mult([-1,c],[d])    
  4 Le([d,a],[])        [L16,2,3,4][L17,1,2,3,4]
\end{lstlisting}
Theorem T25 proves the converse statement that if $-c \leq a \leq c$ then $|a| \leq c$.
The two operand programs that result from disjunction splitting of $Abs([a,0,-1],[b])$ that appears in the premise of T25 are dealt with by lemmas L16 and L18.
Note that this is an example where a lemma, in this case L16, has already been employed by a proof of an earlier theorem.
\begin{lstlisting}
Lemma L18.
[ [ Le([a,c],[]), Aid([a],[b]) ], Le([b,c],[]) ]

Proof.
  1 Le([a,c],[])        
  2 Aid([a],[b])        
  3 Eq([b,a],[])        [A7,2]
  4 Le([b,c],[])        [L5,3,1]

Theorem T25.
[ [ Mult([-1,c],[d]), Le([a,c],[]), Le([d,a],[]), Abs([a,0,-1],[b]) ],
 Le([b,c],[]) ]

Proof.
  1 Mult([-1,c],[d])    
  2 Le([a,c],[])        
  3 Le([d,a],[])        
  4 Abs([a,0,-1],[b])*  
  5 Le([b,c],[])        [L16,1,3,5][L18,2,5]
\end{lstlisting}
When combined, theorems T26 and T27, state that $-|a| \leq a \leq |a|$.
No disjunction splitting is required in the proofs of T26 and T27. 
\begin{lstlisting}
Theorem T26.
[ [ Abs([a,0,-1],[b]) ], Le([a,b],[]) ]

Proof.
  1 Abs([a,0,-1],[b])   
  2 Int([b],[])         [A2,1]
  3 Eq([b,b],[])        [A5,2]
  4 Le([b,b],[])        [A36,3]
  5 Le([a,b],[])        [T23,1,4]

Theorem T27.
[ [ Abs([a,0,-1],[b]), Mult([-1,b],[c]) ], Le([c,a],[]) ]

Proof.
  1 Abs([a,0,-1],[b])   
  2 Mult([-1,b],[c])    
  3 Int([b],[])         [A2,1]
  4 Eq([b,b],[])        [A5,3]
  5 Le([b,b],[])        [A36,4]
  6 Le([c,a],[])        [T24,1,5,2]
\end{lstlisting}
Theorem T28 is the triangle inequality on $\mb{I}$.
It essentially states that if $|x|+|y|$ and $|x+y|$ exist, i.e. are type $\mb{I}$, then $|x+y| \leq |x|+|y|$.
No disjunction splitting is required.
\begin{lstlisting}
Theorem T28.
[ [ Abs([x,0,-1],[u]), Abs([y,0,-1],[v]), Add([u,v],[w]), Add([x,y],[z]),
 Abs([z,0,-1],[p]) ], Le([p,w],[]) ]

Proof.
  1 Abs([x,0,-1],[u])   
  2 Abs([y,0,-1],[v])   
  3 Add([u,v],[w])      
  4 Add([x,y],[z])      
  5 Abs([z,0,-1],[p])   
  6 Le([x,u],[])        [T26,1]
  7 Le([y,v],[])        [T26,2]
  8 Le([z,w],[])        [A34,6,7,4,3]
  9 Int([u],[])         [A2,1]
 10 Mult([-1,u],[a])    [A14,9]
 11 Int([v],[])         [A2,2]
 12 Mult([-1,v],[b])    [A14,11]
 13 Le([a,x],[])        [T27,1,10]
 14 Le([b,y],[])        [T27,2,12]
 15 Int([w],[])         [A2,3]
 16 Mult([-1,w],[c])    [A14,15]
 17 Add([a,b],[d])      [A23,3,16,10,12]
 18 Eq([d,c],[])        [A25,3,16,10,12,17]
 19 Le([d,z],[])        [A34,13,14,17,4]
 20 Int([z],[])         [A2,4]
 21 Eq([z,z],[])        [A5,20]
 22 Le([c,z],[])        [A3,19,18,21]
 23 Le([p,w],[])        [T25,16,8,22,5]
\end{lstlisting}

\section{Consistency.}

From a strict formalist point of view the semantics of statements in a formal system is less of a concern than that of consistency.
In a computer environment a formal statement is expressed in the form of a program whose functionality is largely well defined.
In this context the semantics of formal statements becomes less ambiguous. 
Attaching a truth value to a program as a formal statement can simply be associated with its computability.

Mainstream mathematicians have a more relaxed attitude than their constructivist counterparts regarding the need to establish the truth of a premise before a proof is derived.
The aim of a proof in classical logic is to establish a conclusion that is understood to be true if the premise is true.
This feature of classical logic is reflected in our definition of a computable program extension.
However, it would not be correct to conclude that our formal system is contained fully within classical logic.
It shares features common to both classical and intuitionistic logic. 

Mathematicians often appeal to the law of the excluded middle by starting with a premise that they believe to be false and proceed to derive a proof that leads to a contradiction.
Leading up to the contradiction they obtain formal statements that are derived from axioms and previously derived theorems.
From the point of view of constructivism this is unacceptable because in constructive logic it is meaningless to derive statements from a false premise.

It is important to note that in our formal system derivations based on axioms and theorems of falsity are not equivalent to the classical method of proof by contradiction.
Proofs leading to a conclusion of falsity are based on a higher order type assignment and not a contradiction of the premise.
Furthermore, there is no feature built into \emph{VPC} that calls upon the law of the excluded middle.
Derivations leading up to conclusions of falsity are simply aimed at detecting programs that are not computable for any value assigned input. 
Detecting false programs is of particular interest when dealing with disjunctions through an appeal to the disjunction contraction rules.
This involves a contraction back onto the main proof containing the disjunction only after proofs of all operand programs have been completed. 

\vspace{5mm}
\noindent
\textbf{Computational empiricism.}
Recall that we have constructed our formal system under the constraints defined by the machine specific parameters $K,L$ and $M$ that reflect the finite resources of a machine environment.
So far we have considered only very simple programs that do not involve iterations so the issue of the halting problem appears to be remote.
Having accepted to work within a framework of a formal system constrained by machine specific parameters it would appear reasonable to impose yet another constraint that a program will be deemed to be computable if it halts without encountering an execution error within a prescribed time frame.
We may include amongst the parameters $K,L$ and $M$ the new parameter
\be
\begin{array}{ll}
T & \text{maximum execution time for a single program.} \\
\end{array}
\ee
The parameter $T$ depends largely on the machine processors in combination with a somewhat arbitrary choice of the user.

In such a computer environment one may establish the computability of a program by simply executing that program for the specific value assigned input and observing whether it returns an output or halts with an execution error.
We shall call this procedure computational empiricism.

\vspace{5mm}
\noindent
\textbf{Revised axiomatic method.}
The questions regarding the foundations of mathematics are well known and remain a topic of serious debate.
Rather than attack this problem head on we may seek a path around it.  
One approach is to accept a less ambitious form of inquiry that is more akin to that found through the self improving recursive process of the scientific method.
Consequently the axiomatic method is weakened to incorporate some procedures that may be empirical.
This may not be acceptable to the purist but it may be argued that it better reflects the trial and error process in which mathematics is actually conducted in the real world.
First of all one concedes to the notion that, like postulates in science, laying down a collection of axioms to define a specific theory may be a tentative process that is subject to modification.
It is through such a concession that an iterative mechanism is required for continuous reevaluation and self correction.

One initiates a proof mining activity by first laying down a collection of axioms.
By applying these axioms in conjunction with the construction rules, proofs are derived from which theorems are extracted as irreducible computable program extensions. 
It can be expected that there will be irreducible computable program extensions that are missed by this process because no derivation exists for such statements under the current collection of axioms.
Irreducible computable program extensions that cannot be derived are by definition axioms.
If by some means outside of the proof mining activity a new axiom is found it can simply be appended to the current collection of axioms.
In this way the theory under investigation is built up with increasing scope of the proof mining process. 

\vspace{5mm}
\noindent
\textbf{Identifying new axioms.}
The actual task of identifying new axioms is an activity that lies outside of the formal system in which they are employed.
At this stage such an activity is largely a human enterprise but it is worthwhile to speculate that automation may be possible.
It is difficult to envisage a procedure of identifying axioms that can avoid some kind of empirical process.
This may involve a mechanism of extensive testing of a statement with respect to a large range of prescribed value assigned inputs through computational empiricism in combination with confidence valuation through statistical analysis.   
This would be an activity that would be conducted in parallel to the main activity of generating proofs.
Since our formal system is constrained under the machine specific parameters $K,L,M$ and $T$, we can expect that the empirical procedure just described may identify new axioms that are machine specific.
In a larger realm of investigation the machine specific parameters become variables that enter the self correcting recursive procedure.

\vspace{5mm}
\noindent
\textbf{Consistency.}
In conventional theories of logic consistency is defined in terms of formal statements and their negations.
Here we have not made much use of negations so consistency of our formal system requires a new definition.

Suppose that by derivation it has been established that $s:\mb{P}_{cpe}(p,c)$.
Suppose further that by computational empiricism the program $p$ is found to be computable for a given assigned input.
Consistency will be violated if by means of inference or computational empiricism it is found that $s$ is not computable for the same value assigned input.

In our recursive self improving procedure is included the following two actions that run parallel to the proof mining activity.

\begin{itemize}

\item Axioms are assumed to be irreducible computable program extensions until such time that they are found to lead to an inconsistency.
Inconsistencies can be detected either by derivation or through computational empiricism.
When this occurs the offending programs that are stored as axioms are removed from storage along with all theorems whose derivations relied upon them.

\item If a derivation or proof is later found for an axiom then it is accessed in the file \emph{axiom.dat} and relabelled as a theorem.

\end{itemize}

\vspace{5mm}
\noindent
\textbf{Premature derivation halting.}
Another form of inconsistency may occur if by inference we obtain $P(x,y):\mb{P}_{false}$ and it is later found by computational empiricism that there exists an assigned valued input such that the program $P(x,y)$ is computable.  
A point of caution here is the lack of certainty that we have detected all programs that are of type $\mb{P}_{false}$.
As a result we might extract theorems from derivations that have been halted prematurely with conclusions that do not state the falsity of their premises.
Thus we may have theorems stored in \emph{axiom.dat} that will never be computable.
However, such theorems are benign in the sense that any proof construction starting from a premise program that is computable will never access such theorems.
Once a proof of a premise program has been found with a conclusion of falsity it is stored in \emph{axiom.dat} as a theorem of falsity.
A search can then be conducted of all axioms and theorems currently stored in \emph{axiom.dat} whose premise programs contain, as a sublist, the premise program associated with the theorem of falsity.
When these are identified they are simply removed from storage along with all theorems whose derivations depended on those programs that were stored as axioms/theorems.  

\section{Concluding Remarks.}

Many theorem proving software available today make considerable effort to provide an interface that allows the user to interact in the more familiar language of mainstream mathematics.
In developing \emph{VPC} very little effort has been invested in this area.
This is a choice that is deliberate and is made to encourage familiarity with a language where formal statements are expressed in the form of functional programs.
There is a need to emphasize the importance of representing programs explicitly in terms of the variable names of their I/O lists and the crucial role that the variable names play in governing the processes of program manipulations and derivations.    

Included in future developments of the \emph{VPC} project is the formalization of the procedures discussed in the previous section.
Along with this will be a detailed examination of the current construction rules.
Generalization of the rules can be sought by regarding the construction rules as the axioms of a theory for the construction of programs as proofs and derived within the formal system itself.
So far \emph{VPC} has been applied to some basic results of arithmetic on $\mb{I}$ although it is understood that the methods on which \emph{VPC} is based can be applied to other theories.
However, the primary emphasis of the package is to remain as a tool focused on machine arithmetic and in particular the validation of computer models that are employed in real world applications.

\newpage

\begin{flushleft}

\appendix{\textbf{APPENDIX.}}

\section{Atomic integer programs.}

\vspace{5mm}
\textbf{Check type integer.}

\textbf{\textit{Syntax.}} $Int([a],[~])$.

\textbf{\textit{Program Type.}} $\mb{P}_{type}$.

\textbf{\textit{Type checks.}} $a:\mb{I}$.

\textbf{\textit{Description.}} $Int$ checks that the value assigned to $a$ has type $\mb{I}$.
$Int$ halts with an execution error if there is a type violation.

\vspace{5mm}
\textbf{Less than.}

\textbf{\textit{Syntax.}} $Lt([a,b],[~])$.

\textbf{\textit{Program Type.}} $\mb{P}_{type}$.

\textbf{\textit{Type checks.}} $a:\mb{I}$, $b:\mb{I}$, $a<b$.

\textbf{\textit{Description.}} 
$Lt$ first checks that the values assigned to $a$ and $b$ are type $\mb{I}$.
It then checks that $a<b$.
$Lt$ halts with an execution error if there is a type violation.
Type violation includes the case where the assigned value of $a$ fails to be less than the assigned value of $b$.

\vspace{5mm}
\textbf{Numerical equality.}

\textbf{\textit{Syntax.}} $Eq([a,b],[~])$.

\textbf{\textit{Program Type.}} $\mb{P}_{type}$.

\textbf{\textit{Type checks.}} $a:\mb{I}$, $b:\mb{I}$, $a=b$.

\textbf{\textit{Description.}} 
$Eq$ first checks that the values assigned to $a$ and $b$ are type $\mb{I}$.
It then checks that $a=b$.
Here equality is in the sense of assigned values.
$Eq$ halts with an execution error if there is a type violation.
Type violation includes the case where the value assigned to $a$ fails to be equal to the value assigned to $b$.

\vspace{5mm}
\textbf{Identity assignment.}

\textbf{\textit{Syntax.}} $Aid([a],[b])$.

\textbf{\textit{Program Type.}} $\mb{P}_{assign}$.

\textbf{\textit{Type checks.}} $a:\mb{I}$, $b:\mb{I}$.

\textbf{\textit{Assignment function.}} $b:=a$.

\textbf{\textit{Description.}} $Aid$ first checks that the value assigned to $a$ is type $\mb{I}$.
It then assigns to $b$ the value assigned to $a$, i.e. $b:=a$.
If $a:\mb{I}$ then the type check $b:\mb{I}$ is never violated.
$Aid$ returns the value $b$ as output provided that there are no type violations.
Otherwise it halts with an execution error.

\vspace{5mm}
\textbf{Addition.}

\textbf{\textit{Syntax.}} $Add([a,b],[c])$.

\textbf{\textit{Program Type.}} $\mb{P}_{assign}$.

\textbf{\textit{Type checks.}} $a:\mb{I}$, $b:\mb{I}$, $c:\mb{I}$.

\textbf{\textit{Assignment function.}} $c:=a+b$.

\textbf{\textit{Description.}}
$Add$ first checks that the values assigned to $a$ and $b$ are type $\mb{I}$.
It then attempts to assign to $c$ the sum of $a$ and $b$, i.e. $c:=a+b$ provided that $c:\mb{I}$.
This may fail if the sum $a+b$ is not contained within $\mb{I}$.
$Add$ returns the value $c$ as output provided that there are no type violations.
Otherwise it halts with an execution error.

\vspace{5mm}
\textbf{Multiplication.}

\textbf{\textit{Syntax.}} $Mult([a,b],[c])$.

\textbf{\textit{Program Type.}} $\mb{P}_{assign}$.

\textbf{\textit{Type checks.}} $a:\mb{I}$, $b:\mb{I}$, $c:\mb{I}$.

\textbf{\textit{Assignment function.}} $c:=a*b$.

\textbf{\textit{Description.}}
$Mult$ first checks that the values assigned to $a$ and $b$ are type $\mb{I}$.
It then attempts to assign to $c$ the product of $a$ and $b$, i.e. $c:=a*b$ provided that $c:\mb{I}$.
This may fail if the product $a*b$ is not contained within $\mb{I}$.
$Add$ returns the value $c$ as output provided that there are no type violations.
Otherwise it halts with an execution error.

\vspace{5mm}
\textbf{Division.}

\textbf{\textit{Syntax.}} $Div([a,b],[c])$.

\textbf{\textit{Program Type.}} $\mb{P}_{assign}$.

\textbf{\textit{Type checks.}} $a:\mb{I}$, $b:\mb{I}$, $c:\mb{I}$.

\textbf{\textit{Assignment function.}} $c:=a/b$.

\textbf{\textit{Description.}}
$Div$ first checks that the values assigned to $a$ and $b$ are type $\mb{I}$.
It then attempts to assign to $c$ the value of $a$ divided by $b$, i.e. $c:=a/b$ provided that $c:\mb{I}$.
This may fail if $b=0$ or if $b$ is not an integer multiple of $a$.
$Div$ returns the value $c$ as output provided that there are no type violations.
Otherwise it halts with an execution error.

\section{Atomic higher order programs.}

\vspace{5mm}
\textbf{Check type program.}

\textbf{\textit{Syntax.}} $Prog([p],[~])$.

\textbf{\textit{Program Type.}} $\mb{P}_{type}$.

\textbf{\textit{Type checks.}} $p:\mb{P}$.

\textbf{\textit{Description.}} $Prog$ checks that $p:\mb{P}$, i.e. $Prog$ recognizes $p$ as a string and checks that it does not violate any of the structural conditions stated in Definition 1.
$Prog$ halts with an execution error if there is a type violation.

\vspace{5mm}
\textbf{Check program equivalence.}

\textbf{\textit{Syntax.}} $Equiv([p,q],[~])$.

\textbf{\textit{Program Type.}} $\mb{P}_{type}$.

\textbf{\textit{Type checks.}} $p:\mb{P}$, $q:\mb{P}$, $p \equiv q$.

\textbf{\textit{Description.}} $Equiv$ first checks that $p:\mb{P}$ and $q:\mb{P}$.
It then checks that $p$ and $q$ are program equivalent, i.e. $p \equiv q$.
$Equiv$ will examine the program structures of $p$ and $q$ and identify whether any one of the properties leading to program equivalence as itemized in Definition 10 is satisfied.
$Equiv$ halts with an execution error if there is a type violation.
Type violation includes the case that $p$ and $q$ fail to be program equivalent.

\vspace{5mm}
\textbf{Check I/O equivalence.}

\textbf{\textit{Syntax.}} $Eqio([p,q],[~])$.

\textbf{\textit{Program Type.}} $\mb{P}_{type}$.

\textbf{\textit{Type checks.}} $p:\mb{P}$, $q:\mb{P}$, $p \thicksim q$.

\textbf{\textit{Description.}} $Eqio$ first checks that $p:\mb{P}$ and $q:\mb{P}$.
It then checks that $p$ is I/O equivalent to $q$, i.e. $p \thicksim q$.
$Eqio$ halts with an execution error if there is a type violation.
Type violation includes the case that $p$ fails to be I/O equivalent to $q$. 

\vspace{5mm}
\textbf{Check program sublist.}

\textbf{\textit{Syntax.}} $Sub([q,p],[~])$.

\textbf{\textit{Program Type.}} $\mb{P}_{type}$.

\textbf{\textit{Type checks.}} $p:\mb{P}$, $q:\mb{P}$, $q \subseteqq p$.

\textbf{\textit{Description.}} $Sub$ first checks that $p:\mb{P}$ and $q:\mb{P}$ and then checks that $q$ is a sublist of $p$, i.e. $q \subseteqq p$.
$Sub$ halts with an execution error if there is a type violation.
Type violation includes the case that $q$ is not a sublist of $p$.

\vspace{5mm}
\textbf{Check type computable program extension.}

\textbf{\textit{Syntax.}} $Cpe([p,c],[~])$.

\textbf{\textit{Program Type.}} $\mb{P}_{type}$.

\textbf{\textit{Type checks.}} $p:\mb{P}$, $c:\mb{P}$, $[p,c]:\mb{P}_{cpe}(p,c)$.

\textbf{\textit{Description.}} $Cpe$ checks that $p:\mb{P}$, $c:\mb{P}$ and $[p,c]:\mb{P}_{cpe}(p,c)$.
Axioms and theorems stored in the file \emph{axiom.dat} are automatically assigned the type $\mb{P}_{cpe}$.
Otherwise a program acquires the type $\mb{P}_{cpe}$ through an assignment by way of inference via the construction rules.
$Cpe$ halts with an execution error if there is a type violation.

\vspace{5mm}
\textbf{False program type check.}

\textbf{\textit{Syntax:}} $False([p],[~])$.

\textbf{\textit{Program Type:}} $\mb{P}_{type}$.

\textbf{\textit{Type checks:}} $p:\mb{P}_{false}$.

\textbf{\textit{Description:}} $False$ checks that $p:\mb{P}_{false}$.
A program acquires the type $\mb{P}_{false}$ through an assignment by way of inference via the construction rules.
$False$ halts with an execution error if there is a type violation.

\vspace{5mm}
\textbf{Computable program extension type assignment.}

\textbf{\textit{Syntax.}} $Acpe([p,c],[~])$.

\textbf{\textit{Program Type.}} $\mb{P}_{tassign}$.

\textbf{\textit{Type checks.}} $p:\mb{P}$, $c:\mb{P}$, $[p,c]:\mb{P}$.

\textbf{\textit{Description.}} $Acpe$ first checks that $p:\mb{P}$, $c:\mb{P}$ and $[p,c]:\mb{P}$.
If there are no type violations $Acpe$ then makes the assignment of subtype $[p,c]::\mb{P}_{cpe}(p,c)$.
$Acpe$ halts with an execution error if there is a type violation, i.e. any one of $p$, $c$ and $[p,c]$ is not of type $\mb{P}$.

\vspace{5mm}
\textbf{False program type assignment.}

\textbf{\textit{Syntax:}} $Afalse([p],[~])$.

\textbf{\textit{Program Type:}} $\mb{P}_{tassign}$.

\textbf{\textit{Type checks:}} $p:\mb{P}$.

\textbf{\textit{Description:}} $Afalse$ checks that $p:\mb{P}$.
If there is no type violation $Afalse$ then makes the assignment of subtype $p::\mb{P}_{false}$.
$Afalse$ halts with an execution error if there is a type violation, i.e. $p$ is not of type $\mb{P}$.

\vspace{5mm}
\textbf{Program list concatenation.}

\textbf{\textit{Syntax.}} $Conc([p,q],[r])$.

\textbf{\textit{Program Type.}} $\mb{P}_{assign}$.

\textbf{\textit{Type checks.}} $p:\mb{P}$, $q:\mb{P}$, $r:\mb{P}$.

\textbf{\textit{Assignment function.}} $r:=[p,q]$.

\textbf{\textit{Description.}} $Conc$ first checks that $p:\mb{P}$ and $q:\mb{P}$.
If successful $Conc$ then attempts to assign to $r$ the program concatenation of $p$ and $q$, i.e. $r:=[p,q]$ provided that $r:\mb{P}$.
This may fail if for instance the conditions $y_p \bigcap y_q =[~]$ and $x_p \bigcap y_q =[~]$ are not satisfied.
$Conc$ halts with an execution error if there is a type violation.

\vspace{5mm}
\textbf{Program disjunction.}

\textbf{\textit{Syntax:}} $Disj([p,q],[r])$.

\textbf{\textit{Program Type:}} $\mb{P}_{assign}$.

\textbf{\textit{Type checks:}} $p:\mb{P}$, $q:\mb{P}$, $r:\mb{P}$.

\textbf{\textit{Assignment function.}} $r:=p~|~q$.

\textbf{\textit{Description:}} $Disj$ first checks that $p:\mb{P}$ and $q:\mb{P}$ and then attempts to assign to $r$ the disjunction of the programs $p$ and $q$, i.e. $r:=p~|~q$ provided that $r:\mb{P}$.
$Disj$ halts with an execution error if there is a type violation.

\end{flushleft}

\end{document}